\documentclass[acmsmall,nonacm,screen]{acmart}
\usepackage{tikz}
\usepackage{ragged2e}
\usepackage{array}
\usepackage{threeparttable}
\usepackage{subcaption}
\usepackage{tabularx, siunitx, hanging}
\usepackage[nameinlink,capitalize]{cleveref}
\usetikzlibrary{arrows.meta,positioning}

\AtBeginDocument{%
  }

\setcopyright{none}
\acmYear{2023}
\acmDOI{}

\begin{document}

\title[Blockchain Software Health]{Towards a Structural Equation Model of Open Source Blockchain Software Health}
\author{Jeff Nijsse}
\email{jeff.nijsse@aut.ac.nz}
\orcid{0000-0002-4199-7201}
\author{Alan Litchfield}
\email{alan@alphabyte.co.nz}
\orcid{0000-0002-3876-0940}
\affiliation{%
  \institution{Auckland University of Technology}
  \streetaddress{Private Bag 92006}
  \city{Auckland}
  \country{New Zealand}
  \postcode{1142}
}
\renewcommand{\shortauthors}{Nijsse and Litchfield}
\begin{abstract}

The widespread use of GitHub among software developers as a communal platform for coordinating software development has led to an abundant supply of publicly accessible data. Ever since the inception of Bitcoin, blockchain teams have incorporated the concept of open source code as a fundamental principle, thus making the majority of blockchain-based projects' code and version control data available for analysis.

We define health in open source software projects to be a combination of the concepts of sustainability, robustness, and niche occupation. Sustainability is further divided into interest and engagement. This work uses exploratory factor analysis to identify latent constructs that are representative of general public \textit{Interest} or popularity in software, and software \textit{Robustness} within open source blockchain projects. We find that \textit{Interest} is a combination of stars, forks, and text mentions in the GitHub repository, while a second factor for \textit{Robustness} is composed of a criticality score, time since last updated, numerical rank, and geographic distribution. Cross validation of the dataset is carried out with good support for the model.

A structural model of software health is proposed such that general interest positively influences developer engagement, which, in turn, positively predicts software robustness. The implications of structural equation modelling in the context of software engineering and next steps are discussed.

\end{abstract}




\maketitle
\section{Introduction}\label{sec:intro}
Software health is a multifaceted and elusive concept, drawing parallels with biological and environmental health as well as business health and their respective ecosystems. Studies in software health have explored various perspectives, from examining natural ecosystems~\cite{Costanza1992,Schaeffer1988} to contributor motivation~\cite{Hars2001,Bosu2019}, software communities~\cite{Goggins2021,Shi2021} and their wider ecosystems~\cite{vandenBerk2010,Goeminne2013} in order to derive representative models. Generally, these models refer to the overall well-being of a software system, which encompasses its performance, reliability, maintainability, and other related factors. Just like human health, software health is a critical aspect that affects the functionality and longevity of software systems.

Understanding and maintaining software health is important for the success of software systems and the satisfaction of software developers and end-users. By measuring and improving software health, organizations can ensure the long-term viability of their software systems and avoid costly downtime and maintenance issues. When software systems are healthy, developers can work more efficiently, productively, and with fewer errors. This, in turn, can lead to increased satisfaction and morale, which can have a positive impact on developer retention and recruitment. Similarly, end-users are more likely to be satisfied with a software product that is healthy and responsive to their needs, which can lead to increased user adoption and loyalty.

However, measuring the health of software can be challenging due to several factors. For instance, there is a lack of consensus on a definition of exactly what is meant by "health," and different stakeholders such as developers and end-users may have different perspectives. Additionally, version control data may be limited for proprietary software and corporations, combined with a dearth of user-friendly tools to assist with measurement. Finally, it can be difficult to identify constructs such as \textit{robustness} and \textit{engagement} that are subjectively critical to health, but objectively hard to measure.

GitHub is the largest open-source software (OSS) community with over 254 million repositories by 73 million developers~\cite{GitHub2021, Gousios2016, Hu2016}. The platform has been pivotal in the rise of cryptocurrencies and blockchain projects, with 98.6\% of the top 419 public repositories in this field hosted on GitHub~\cite{Nijsse2023}. The open-source ethos of decentralization has played a significant role in the rapid innovation and iteration seen in this field. Unfortunately, with technical innovation in blockchain can come social harm. 2022 was the worst year on record for the crypto industry with over \$3 billion USD hacked, stolen, or lost from crypto-related projects and exchanges~\cite{Schwartz2022}. Therefore, understanding and maintaining software health is critical to prevent such losses and ensure the long-term viability of blockchain software systems.

By analysing the health of OSS in the scope of blockchain projects, this work contributes to the larger theme of software health and highlights the importance of measuring and maintaining software health in various domains, including emerging technologies like blockchain. Through this approach, researchers and stakeholders can develop more comprehensive models and strategies to ensure software health.

The objective of this work is to develop a model of software health in OSS blockchain projects using publicly available data from GitHub. Our hypothesis is that identifiable factors that contribute to the overall health of a software project can be measured and analysed. To achieve this, the statistical methods exploratory factor analysis (EFA) and structural equation modelling (SEM) are utilized to derive factors that inform the model of software health. With this model, researchers and practitioners can identify areas of strength and weakness within blockchain software to maintain high-quality, reliable, and efficient software systems.


\subsection{Research Questions}\label{sec:RQs}
Given that metrics can be derived from publicly available version control software, the following research questions are investigated with respect to software health in the blockchain domain.

\begin{enumerate}
	\item[\textbf{RQ1:}] What facets make up a high-level definition of software health, without using or defining specific metrics?
	\item[\textbf{RQ2:}] Given that software robustness is a contributing component of health, what are the factors that contribute to software robustness?
	\item[\textbf{RQ3:}] How does community interest or, general popularity, fit into the definition of health and what are the component metrics?
	\item[\textbf{RQ4:}] What is the nature of the relationship between components in the definition of software health?
\end{enumerate}

In order to address the research questions posed, this study utilizes a combination of methods including a literature review to inform a definition of health, and applying a framework for analysing open source data via exploratory factor analysis and structural equation modelling. Through these methods, the study aims to derive specific indicator metrics that contribute to software health. The literature review provides a foundation for the study, while the framework for health operationalization guides the collection and analysis of open source data. The use of EFA and SEM allows for the identification of underlying factors and the exploration of relationships between variables. Overall, this approach provides a rigorous and systematic method for investigating the research questions at hand.

\subsection{Contributions to the Field}\label{sec:contributions}
The results of this study have several contributions to the broader scientific fields of information systems and software engineering.

\begin{enumerate}
	\item Enhancing the existing literature on software health by providing a quantifiable definition of software health based on factors identified through exploratory factor analysis and structural equation modelling. This is an important contribution as a clear and measurable definition of software health can help researchers and practitioners to evaluate and compare software systems, and make data-driven decisions related to software development and maintenance.

	\item A novel application of latent factor analysis to identify the factors that contribute to both community interest and software robustness in open source blockchain software projects. This approach provides a more comprehensive understanding of how these factors influence software health.

	\item The proposed structural equation model offers a new way of thinking about software health and its various components, which can help project managers and developers identify areas for improvement and optimize resource allocation. This model represents an advancement in the field of software health analysis and provides the groundwork for future research.

	\item In addition, the study makes a valuable contribution to the field by providing a publicly available dataset, which can be used by researchers and practitioners to further investigate software health in open source blockchain software projects. This dataset includes the entire set of metrics related to developer engagement, software robustness, and community interest; factors that influence software health model found in the study.
\end{enumerate}

The remainder of the article is structured as follows: In \cref{sec:health} the background literature and related work in the fields of software health are discussed. This provides the necessary context for understanding the research questions and the contributions of the study. The scope is narrowed to blockchain software in \cref{sec:blockchain-health}. \cref{sec:methods} presents research methodology, which includes a detailed description of the framework used to analyse open source data, the statistical methods employed in the study, and the data collection details are provided, including the sources of data, the criteria used for selection, and the steps taken to clean and preprocess the data. Results are presented in \cref{sec:results}, including the findings from the latent factor analysis and the structural equation model. Implications of the results are discussed in \cref{sec:discussion}, including the practical applications of the findings for stakeholders, the limitations of the research, and provides suggestions for future work in this area. Finally, the conclusion (\cref{sec:conclusion}) summarizes the main results and contributions of the paper.

\section{Software Health}\label{sec:health}
Before we get to the background literature and defining software health, we will briefly mention open source software and what it means in the present context.
\subsection{Open Source Software}\label{sec:OSS}
The term `open source' emerged from free and open source software (FOSS) in 1998 when Netscape chose to release its web browser Netscape Communicator as FOSS\footnote{Netscape’s move to open source their browser was radical and so to prepare enterprises for the news they opted to use the term `open source’ to be more business friendly than the ambiguous term `free’.}~\cite{Gonzalez2021oss}. Bruce Parens was at Netscape during the code release and worked extensively to write a new definition of open source\footnote{Parens also formed the Open Source Initiative (OSI) to steward the project and develop the license. This was motivated by Eric Raymond publishing \textit{The Cathedral \& the Bazaar} seven months earlier.}, the essence of which is that  OSS (i) is free to distribute without royalties, (ii) has published source code, (iii) and can be modified by anyone with few conditions~\cite{Perens1999}. The more verbose ten-point version can be found on the opensource.org website\footnote{\url{https://opensource.org/osd.html}}.

In the context of this study there must be source code and version tracking information available for analysis. See \cref{sec:dataset} for data collection details. It is noted that there are different types of open source licenses, and this is not of concern to the present work. We do not make any assumptions or exclusions based on a projects’ software license. For example Bitcoin was published under the MIT/X11 license, while Ethereum's source code is licensed under the GNU Lesser GPL. Both licenses allow for copying (forking) and republishing of code.

\subsection{Ecosystem Health as a Metaphor for Software Health}\label{sec:health-definition}
Defining health in the context of software, and further blockchain software, will first benefit from a view of health as seen in the life sciences. The health of natural ecosystems and their components, such as soil, water, flora, and fauna, is a pressing concern for the entire biosphere. As we seek to understand what constitutes natural ecosystem health, we may find it useful to draw on metaphors, even those from human medicine~\cite{Rapport1989, Hartigh2013}. Metaphor has a legitimate place in science as it can stimulate associations between seemingly unrelated phenomena and highlight their structural identity~\cite{Rapoport1983}.

The ecological metaphor has been used extensively to relate natural ecosystems to both business ecosystems~\cite{Moore1993, Iansiti2004, Hartigh2013}, and software ecosystems~\cite{Dhungana2010, Chengalur2010, Goeminne2013, Manikas2013, Jansen2014}, and allows parallels to be drawn on the basis of health. Both natural and software ecosystems are composed of interrelated components, such as species in natural ecosystems and projects in software ecosystems that exist in a competitive environment. Both ecosystems rely on biodiversity in order to thrive, and an underlying principle in both is that of adaptation and evolution of the components within the system in order to ensure its survival and continued success. By learning from the relationships in natural ecosystems, we can identify factors that are crucial to the sustainability and overall health in software.


A conceptual map illustrating the different ways the idea of health is defined across natural, business, software, and open source ecosystems is presented in \autoref{tab:health}. This map shows the key terms used to describe health in each ecosystem and categorizes them into three groups: sustainability, robustness, and niche fit. The terms were identified through a comparative analysis of ecosystem-specific literature, and provide insights into the challenges present when defining health.

\begin{table}[h]
	\centering
	\begin{threeparttable}
	\caption{Conceptual map illustrating the different ways in which the concept of health is defined across natural, business, software, and open source ecosystems. This map shows the key terms used to describe health in each ecosystem and categorizes them into three groups: sustainability, robustness, and niche fit.}
	\label{tab:health}{\small
	\begin{tabularx}{\linewidth}{
		    @{}l
		    >{\hangpara{3mm}{1}\raggedright\arraybackslash}X
			p{3cm}
			>{\raggedright\arraybackslash\hangafter=1\hangindent=3mm}p{2.5cm}
			l
			r
			r@{}
		}
		\toprule
			Ecosystem & Classifier 1 & Classifier 2 & Classifier 3 & & Year & Source\\
		\cmidrule{1-4}\cmidrule{6-7}
	Natural & Productivity & Absence of disease & Diversity & &1988& \cite{Schaeffer1988} \\
		& Sustainability & Integrity; Stress capacity &  & &1989& \cite{Rapport1989} \\
		& Vigour & Resilience & Organization & &'92,'98&\cite{Costanza1992,Rapport1998} \\

		\cmidrule{1-4}\cmidrule{6-7}
	Business & Growth & Profitability & Stable value-creation & & 1993 &\cite{Moore1993} \\
		& Productivity & Robustness & Niche creation & & 2004 &\cite{Iansiti2004} \\
		& Financial health & Centrality; Visibility & Variety of partners&  & 2013  &\cite{Hartigh2013}\\
	\cmidrule{1-4}\cmidrule{6-7}
	SECO\tnote{$\ast$}& Productivity & Robustness & Niche creation&  &2010 & \cite{vandenBerk2010} \\
	    & Productive & Endure & Variable & & 2013 & \cite{Manikas2013} \\
	\cmidrule{1-4}\cmidrule{6-7}
	OSS\tnote{$\dagger$} & Liveness of users/devs & & & & 2007 &\cite{Wahyudin2007} \\
		& Software development & Long-term & & & 2010 &\cite{Chengalur2010} \\
		& Vigour & Resilience & AMI\tnote{$\star$} &&  2012  &\cite{Raja2012}\\
		& Sustainability; Maintenance capacity & Resource health & Network health; Process maturity & & 2014&\cite{Franco2014} \\
		& Healthy community & Healthy commons & & & 2015  &\cite{Naparat2015} \\
		& Community & Code; Resources &&  & 2018 &\cite{Link2018} \\
		& Sustainability & Survivability & & & 2021 &\cite{Goggins2021} \\
		\cmidrule{1-4}\cmidrule{6-7}
	\textit{Concept} & \textbf{\textit{Sustainability}} & \textbf{\textit{Robustness}} & \textbf{\textit{Niche Fit}} &&  &\\
		\bottomrule
	\end{tabularx}}
\begin{tablenotes}[flushleft]\footnotesize 
	\item[$\ast$] software ecosystem
	\item[$\dagger$] open source software
	\item[$\star$] average mutual information
\end{tablenotes}
\end{threeparttable}
\end{table}

Three concepts have been synthesized from the literature and will serve as our definition: Health, in the open source software context is composed of three broad components: sustainability of day-to-day operations, robustness to stress, and niche occupation within the software's  ecosystem.

\subsection{Sustainability}
Sustainability in a natural ecosystem is also referred to as stability~\cite{Costanza1992}, vigour~\cite{Rapport1998}, and productivity~\cite{Schaeffer1988}, and generally refers to the ecosystem's ability to carry out the basic functions necessary for metabolism and growth. Indicators that an ecosystem is functioning include: primary productivity -- how much growth is occurring, the nutrient base available, the diversity of species present, the amount of instability, disease prevalence, diversity of size spectra, and levels of contaminants~\cite{Rapport1989}. The list provides a first draft of metrics biologists can track to asses sustainability in a natural ecosystem.

If a natural ecosystem is an interrelated collection of species, a business ecosystem is an interrelated collection of businesses across industries that are both in competition and cooperation with each other~\cite{Moore1993}. Further, the health of that business ecosystem is an aggregate of the stability needed to be profitable and the stability needed to grow. Sustainability here is the productivity that comes from general tasks employees undertake to maintain business operations~\cite{Iansiti2004}. Just as base metabolic resources are required to keep species in an ecosystem in competition, healthy financial resources can sustain a business ecosystem to provide the opportunity for innovation and growth~\cite{Hartigh2013}.

Drawing from the definition by Jansen et al.~\cite{Jansen2009}, a software ecosystem (SECO) is a group of stakeholders ``functioning as a unit and interacting with a shared market for software and services.'' Software ecosystems as a class of business ecosystems often operate through a common technological platform, such as Apple's iOS, and participate in the affiliated markets, such as Apple's App Store.

Sustainability within a SECO is doing enough of the minimum viable activity to run day-to-day operations. When done well this productivity allows a software business to compete and possibly thrive. When done poorly a lack of productivity will result in losing market share to a competitor. A productive SECO's outputs include software development activities such as writing code, reviewing, and feature implementation~\cite{Manikas2013}. Ensuring the sustainability of the SECO is a complex process that requires significant community effort and resources from the ideation stage to the version release stage~\cite{Negoita2019}. In modern software development, a product is no longer viewed as a static entity after its initial release; rather, there is a constant need for user feedback, bug fixing, and iteration, all of which are activities that contribute to sustainable open source development~\cite{Robles2005, Negoita2019}.

Not only does OSS require financial resources and software management as in business ecosystems, but there is the added component of having the project sustained by the community~\cite{Arantes2011}. Failure in any of these areas can leave a project abandoned and thus sustainability is a base component of OSS development. Thus, any efforts that ensure ongoing day-to-day software development and its related outputs can be viewed as sustainable~\cite{Negoita2019, Ghapanchi2015}.

\begin{figure}[h]
	\centering
	\begin{tikzpicture}[level distance=1.5cm,
		level 1/.style={sibling distance=3cm},
		level 2/.style={sibling distance=3cm}]
		\node {Health}
		child {node {Sustainability}
			child {node [text width=3cm] {\phantom{aaa}Engagement\phantom{aa} (also: Productivity)}}
			child {node [text width=3cm] {\phantom{aaaa}Interest\phantom{aaa} (also: Popularity)}}
		}
		child {node {Robustness}}
		child {node {Niche fit}};
	\end{tikzpicture}
	\caption{Summarizing the language used to define health as composed of sustainability, robustness, and niche fit. Further, sustainability is composed of engagement or productivity and interest or popularity.}
	\Description{A tree diagram summarizing the language used to define health (the root) as composed of sustainability, robustness, and niche fit. Further, sustainability is composed of engagement and interest.}
	\label{fig:health-tree}
\end{figure}
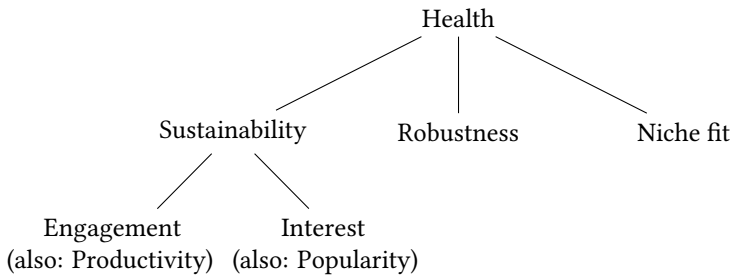

The metrics pertaining to sustainability have been organized in \autoref{tab:metrics-sustainability}. Language variability that is present in analysing the literature show the terms productivity and engagement indicating the same construct. Here we use \textit{Engagement} (as shown in \autoref{fig:health-tree}) to mean any activity that is undertaken to sustain the software project in day-to-day operations. The second sub-classification of sustainability is \textit{General Interest} or popularity. Interest has no parallel in the natural ecosystem literature, but emerges as important to health and success within the software domain in many studies~\cite{Ghapanchi2015, Jansen2014, Saini2020, Wahyudin2007}.


\begin{table}[h]
	\centering
	\begin{threeparttable}
	\caption{Metrics relating to sustainability in the OSS health literature can be split into engagement \& productivity, and interest \& popularity. Often metrics have difference descriptions in the literature, for example the number of downloads can produce a popularity ranking, as can a web ranking.}
	\label{tab:metrics-sustainability}{\small
	\begin{tabularx}{\linewidth}{@{}p{2.8cm}>{\hangpara{3mm}{1}}Xp{2.8cm}@{}}

\multicolumn{3}{l}{\textit{Engagement \& Productivity}}\\
\toprule
		Metric   			& Description and alternative labels 	& Supporting literature   \\
		\midrule
		Bug fix rate    	& How quickly bugs are noted and fixed; also issues or rate of issues opened and closed   & \cite{Crowston2006, Goggins2021, Jansen2014, Negoita2019, Wahyudin2007, Raja2012, Osman2021, Robinson2016, Chengalur2010,  Saini2020} \\
		Comments       		& Volume; including code review, pull request, \& issues-based & \cite{Goggins2021, Jansen2014, Wahyudin2007, Hata2022, Robinson2016, Tamburri2019} \\
		Commits      		&  Count of commits to codebase; also lines of code, method count, token count, project size & \cite{Ghapanchi2015, Goggins2021, Osman2021, Negoita2019, Shaikh2019, Jansen2014}\\
		Contributors 		& Count of contributors, includes developers; also community growth & \cite{Goggins2021, Jansen2014, Osman2021, Saini2020, Tamburri2019, Chengalur2010, Crowston2006, Wang2016}\\
		Pull requests 		& Count of pull requests initialized; also PRs merged, or closed & \cite{Chengalur2010,   Robinson2016, Tamburri2019}\\
		Hours worked\tnote{$\ast$} 		& Productivity measure of contributors' hours spent  & \cite{Goggins2021} \\
		Developer churn\tnote{$\ast$}      & Developers entering and leaving the project/SECO    & \cite{Ghapanchi2015,   Shaikh2019}\\
		Releases\tnote{$\ast$}  			& Count of version releases published &  \cite{Ghapanchi2015,   Raja2012} \\
		Financial resources\tnote{$\ast$}  & Related to business operations   & \cite{Goggins2021,   Jansen2014}\\
\midrule
\multicolumn{3}{l}{\textit{Interest \& Popularity}} \\
\midrule
Dependencies & The number of software dependencies a project has, for example Bitcoin relies on the GCC compiler collection & \cite{Saini2020}                                          \\
		Forks         & Count of number of times the software has been forked & \cite{Jansen2014, Negoita2019, Osman2021, Saini2020}\\
		Rank & Ranking of the project in the broader web, for example number of search engine hits, or Alexa page ranking; also downloads & \cite{Crowston2006, Goggins2021, Jansen2014, Saini2020, Ghapanchi2015, Wahyudin2007}   \\
		Stars         & Count of total stars on GitHub; also tags or watchers & \cite{Negoita2019, Saini2020, Tamburri2019, Osman2021}\\
		\bottomrule
	\end{tabularx}}
\begin{tablenotes}[flushleft]\footnotesize 
	\item[$\ast$]  Metric is considered out of scope and not collected in the present study.
\end{tablenotes}
\end{threeparttable}
\end{table}

Tables~\ref{tab:metrics-sustainability},~\ref{tab:metrics-robustness}, and~\ref{tab:metrics-niche} show the metrics from the literature review, but not all of these are collected and used in the present study. Many of the metrics are lacking a suitable definition, have no hypothesized operationalization using version control information, or are not applicable to the present study of OSS health through publicly available data. Additionally, some metrics concentrate on motivations of individuals that are very difficult to determine without targeted survey data, or are components of a business process such as marketing and financial data that does not apply or is proprietary when limiting the scope to OSS. The selection of specific metrics is a primary goal of this study and discussed further in \cref{sec:methods}~--~\nameref{sec:methods}.

\subsection{Robustness}
Once sustainability has been established the ecosystem can remain productive over time only if it can sustain shocks that threaten its viability. Biodiversity is important to natural ecosystems and helps to enable recovery from shocks. This is through many species performing the same function such as photosynthesis or decomposition, and also by individual species having unique environmental response to threats when compared to their close relatives~\cite{Dhungana2010}. Costanza~\cite{Costanza1992} and Rapport et al.~\cite{Rapport1998} both identify resilience as the ability of an ecosystem to overcome disruption to its local environment.

Environmental factors affect all members of the ecosystem and are generally out of control of the community itself. External risks are more difficult to identify such as misaligned product-market fit, the competitor landscape, technological innovation, and the regulatory and legal landscape~\cite{Chengalur2010}. Many of these dramatic shocks are difficult to quantify and out of scope when considering project level software.

\autoref{tab:metrics-robustness} lists the metrics from the literature contributing to robustness within a SECO. Robustness in software includes demographic factors: age and size of the population, or of the software organization. Both here are positive indicators -- long time contributors are more likely to keep contributing, and long-standing projects have managed to survive likely absorbing prior shocks. Resilience can be in the form of geographic location to avoid some of the environmental risks affecting all participants; and market share information to gauge external validation for the project. Validation from within the community comes from others using, incorporating, and then depending on your software, which can be quantified with a criticality measure~\cite{OSSF2022}.


%
%

\begin{table}[h]
	\begin{threeparttable}
	\caption{Metrics relating to robustness in the OSS health literature. Many robustness measures are difficult to gauge or subjective in nature such as code quality and knowledge creation.}
	\label{tab:metrics-robustness}{\small
	\begin{tabularx}{\linewidth}{@{}p{3.3cm}>{\hangpara{3mm}{1}}Xp{1.4cm}@{}}

\multicolumn{3}{l}{\textit{Robustness}}\\
\toprule
Metric   			& Description and alternative labels 	& Literature   \\
\midrule
    Developer longevity         & Time spent contributing to a single project; also project age            & \cite{Chengalur2010,Goggins2021, Osman2021}   \\
    Geographic distribution     & Global geographic distribution of the contributors  & \cite{Tamburri2019} \\
    Market share       & Ratio of a project's share to the total local ecosystem    & \cite{Jansen2014}  \\
    Project criticality         & Risk associated with project centrality and dependency; also truck factor & \cite{Jansen2014, Tamburri2019, Goggins2021}  \\

    Business metrics\tnote{$\ast$}            & Including management, process development, and systems development; also switching costs   & \cite{Goggins2021, Jansen2014} \\
	End user metrics\tnote{$\ast$}            & Including count, longevity, loyalty, and satisfaction                & \cite{Goggins2021, Wang2016, Jansen2014}\\
	Contributor metrics\tnote{$\ast$} & Including centrality, reputation, satisfaction, cross org participation; also measures of centrality in wider SECO and partnerships  & \cite{Jansen2014, Wahyudin2007}   \\
	Code quality\tnote{$\ast$}                & As relates to code metrics such as cyclomatic complexity             & \cite{Osman2021,Goggins2021}\\
		Knowledge creation\tnote{$\ast$}          & Knowledge added to SECO and artefact creation                        & \cite{Jansen2014}    \\

\bottomrule
\end{tabularx}}
\begin{tablenotes}[flushleft]\footnotesize 
	\item[$\ast$]  Metric is considered out of scope and not collected in the present study.
\end{tablenotes}
\end{threeparttable}
\end{table}

\subsection{Niche Fit}
A third characteristic in the definition of health is that of occupying a relevant niche in the ecosystem. Called organization in the natural ecosystem~\cite{Rapport1998} meaning a species is present at all levels and functions of the aggregate system.

This is paralleled as niche creation in business ecosystems~\cite{Iansiti2004}, and software ecosystems~\cite{Jansen2014}. In areas with ongoing competition among similar products, projects must pivot and identify a unique niche to achieve success. At the software project level it has been suggested that product fit can be quantified by audience niche, programming language niche, and operating system niche~\cite{Chengalur2010}. To have the best chance at occupying a niche, a project may also choose to support multiple natural languages, push applicability to a variety of markets, and be open to various contributor roles~\cite{Jansen2014}. These niche metrics are shown in \autoref{tab:metrics-niche}.


%
\begin{table}[h]
	\caption{Metrics relating to the niche occupancy of a project within an ecosystem in the health literature. Although part of the health literature, all of these metrics are out of scope for collection in the present study.}
	\label{tab:metrics-niche}{\small
	\begin{tabularx}{\linewidth}{@{}p{3.8cm}>{\hangpara{3mm}{1}}Xp{1.4cm}@{}}

	\multicolumn{3}{l}{\textit{Niche Fit}}\\
	\toprule
	Metric   			& Description and alternative labels 	& Literature   \\
	\midrule
		 SECO project variety & Variety in types of projects in the ecosystem demonstrating available   niches & \cite{Jansen2014}    \\
		 Platform variety & Support for a variety of languages allows for new contributors to participate, both natural and programming; also variety in operating systems supported & \cite{Jansen2014, Chengalur2010} \\
		 Market variety       & Cross over applicability of the project to different markets   & \cite{Jansen2014}    \\
		 Contributor types     & Variation in available contributor roles& \cite{Jansen2014}   \\
		 Average mutual information  & A measure of task specialisation and the coordination of specialists & \cite{Raja2012}  \\
		 Niche size           & More member organizations within a niche add legitimacy  & \cite{Goggins2021, Chengalur2010}   \\
		\bottomrule
	\end{tabularx}}
\end{table}
Although Table \cref{tab:metrics-niche} lists items that are reasonably simple to tabulate such as programming language and operating systems, it is not clear how this helps position a project within its local ecosystem niche. Additionally, it is difficult to contextualize a project in the wider landscape when focussing on individual projects and thus niche fit is an aggregate measure that requires a broad view of the whole ecosystem. Therefore, niche related metrics are not within the scope of the study.\\

\subsection{Summary of Software Health}\label{sec:health-summary}
To conclude on open source software health, a high level definition is that health is composed of sustainability, robustness, and the niche occupation of the project as shown in \autoref{fig:health-tree}. This is similar to the health of a natural ecosystem depending on its productivity and growth, resilience, and organisation, respectively.

Within the OSS class of ecosystems, sustainability involves the engagement of developers and other contributors to sustain the ongoing operations necessary to produce software, and also a public interest component where a popular project can attract and retain new talent. The goal of this work is to determine how the metrics available from OSS version control software relate to the definition of health in the scope of blockchain software.

\section{Blockchain Health}\label{sec:blockchain-health}
Bitcoin has inspired a whole new industry in blockchain software development since its release in 2008~\cite{Nakamoto2008}. This has been tracked extensively by the website CoinMarketCap~\footnote{\url{https://coinmarketcap.com}} (CMC) beginning in 2012. Presently, CMC tracks over 20,000 tokens and projects, all that have a blockchain pedigree.

Blockchain software can be defined collectively by the myriad projects listed on CMC including: cryptocurrencies, platforms, protocols, Web3.0 applications, stablecoins, and support libraries of smart contracts. This collection of software is markedly different from the surrounding software industries in the landscape, for example, utilities, databases, web, and mobile.

Blockchain software presents a stronger emphasis on security and reliability than non-blockchain projects~\cite{Bosu2019}. Blockchain-based projects may be more likely to prioritize security and reliability in their development processes. This focus on quality may be driven by the high cost of defects in blockchain software due to direct financial risk of failure, as well as the complex and unique tools used in blockchain development. Additionally, blockchain developers have to work in a more decentralised environment, they skew younger in age, are more educated, and more male than in other software industries~\cite{Bosu2019}.

Bitcoin and newer blockchain projects are highly open-sourced; with the present study finding  69.8\% of the top 600 blockchain projects having publicly visible code repositories on GitHub. This open-sourced, decentralised ethos affects contributor's preferences too, as most projects are started entirely by volunteers playing by tokenised incentives (as opposed to commercial ones~\cite{Smirnova2022}) and these developers will self-select what project to contribute to based on similar values~\cite{Bosu2019}.

Health in blockchain projects and blockchain ecosystems has been studied in a limited manner. One study investigated the health of the Bitcoin ecosystem as defined through the categories: popularity, complexity, activity, and age~\cite{Osman2021}. The authors include code-based metrics within the complexity category which are left out of many other studies on health, however their primary indicators are heuristic-based, chosen without rationale. There are no known studies on health of other blockchain projects such as Ethereum or Solana, or a wider collections of projects or their ecosystems. This absence of research in relation to blockchain software both at the individual developer level and at the software project level is the focus of the present study.

\subsection{Developer Engagement}\label{sec:developer-engagement}
Developer engagement is a component of software sustainability as shown in \autoref{fig:health-tree} and can be described as the day-to-day operations to create and maintain software. In the business sense this is called productivity, and in natural ecosystems involves the necessary metabolic processes for growth.

Developer engagement as a stand alone construct is known to be important~\cite{Poba2019, Fang2008, Shaikh2019, Tamburri2019} but does not itself have a clear index of the components that make up engagement. Software community engagement is a subjective term referring to how people interact and contribute to a project, and includes activities both coding and non-coding related: managing the community, development, documentation, and participating in discussions~\cite{Wang2020}.

\begin{figure}[h]
	\centering
	\includegraphics[trim={0.3cm 0.3cm 0.3cm 0.3cm}, clip, width=0.5\linewidth]{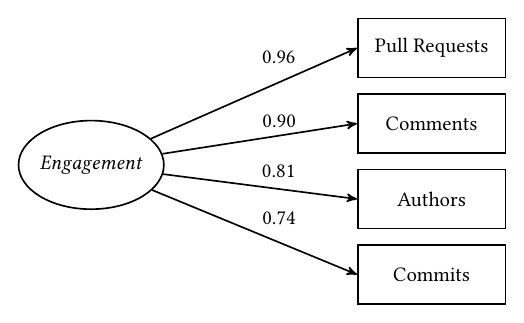}
	\caption{Exploratory factor analysis applied to metrics representing developer engagement. The number of pull requests as a monthly average over the previous three months has the strongest influence on engagement with a factor loading of 0.96.}
	\Description{Exploratory factor analysis diagram showing engagement as a latent factor composed of four indicators: commits, comments, pull-requests, and authors}
	\label{fig:engagement}
\end{figure}

Previous work with the focus on OSS blockchain engagement was completed by the authors finding that the latent factor of developer engagement can be determined with four indicator metrics: commits, comments, pull-requests, and authors~\cite{Nijsse2023}. The present work extends the \textit{Engagement} dimension (see \autoref{fig:engagement}) to include metrics for \textit{Interest} and \textit{Robustness} using similar methods--factor analysis, and new methods--confirmatory factor analysis. The research methodology is discussed presently.

\section{Methods}\label{sec:methods}
The research methodology applies an Open Source Ecosystem Health Operationalization (OSEHO) framework by Jansen~\cite{Jansen2014} modified to assess individual projects within the blockchain ecosystem. This framework is integrated with the work of Goemminne and Mens~\cite{Goeminne2013} on Analysing Open Source Software Ecosystems and involves the steps summarized in \cref{tab:methods}.

\begin{table}[h]
	\centering
	\caption{Research methodology combining Frameworks by Jansen~\cite{Jansen2014} and Goemminne and Mens~\cite{Goeminne2013}.}
	\label{tab:methods}{\small
	\begin{tabular}{lp{0.23\linewidth}p{0.62\linewidth}}
		\toprule
		Step & Action & Description \\
		\midrule
		1 & Set goals & Determine indicator metrics for \textit{Interest} and \textit{Robustness}. \\
		2 & Select ecosystem scope & Limited to blockchain open source software. \\
		3 & Select metrics & As by literature review in \cref{sec:health}: Software Health. \\
		4 & Assess available data & \autoref{tab:efa-metrics} shows the collected metrics. \\
		5 & Data extraction & See \cref{sec:dataset}: Dataset. \\
		6 & Data post-processing & See \cref{sec:processing}: Data Processing. \\
		7 & Statistical analysis & See \cref{sec:results-EFA}: Exploratory Factor Analysis, and \cref{sec:results-CFA}: Structural Equation Modelling. \\
		8 & Reporting & See \cref{sec:results}: Results. \\
		\bottomrule
	\end{tabular}}
\end{table}
First the goal of the study is selected to find the metrics that can form the constructs of \textit{Robustness} and \textit{Interest}. Next, the scope is narrowed to blockchain projects. Step Three is to determine the possible metrics that relate to the concepts through literature review (\cref{sec:health}). Notably, most of these metrics will be out of scope, and so in Step Four the practical selection process is made given resource constraints such as available data and time. These are detailed in \cref{tab:efa-metrics}. Steps Five and Six are to collect the data and prepare the dataset; details of which such as sourcing and cleaning are in \cref{sec:dataset}-\ref{sec:processing}. The analysis is done via statistical methods in \cref{sec:results-EFA} and \cref{sec:results-CFA}. Lastly, reporting of results follows in \cref{sec:results} \nameref{sec:results}.

\subsection{Exploratory Factor Analysis}\label{sec:EFA}
Exploratory factor analysis (EFA) is a multivariate statistical technique used for determining underlying constructs that are present in a dataset composed of a large number of variables. The constructs, or factors, can represent groupings within the data that are hypothesized, or known, but not directly observable~\cite{Clark2018}. In the present context this is the idea of \textit{Robustness} and \textit{Interest} as applied to an open source software project, and the large number of variables are the possible set of metrics identified in Tables~\ref{tab:metrics-sustainability}--\ref{tab:metrics-niche}. \textit{Latent variables} and \textit{factors} are terms used interchangeably and represent inherent characteristics identified by the researcher that do not have a well known associated metric. EFA is common in fields like psychology and economics where participants are given a survey and the results are analysed~\cite{Finch2020} but little work has been done applying the technique to software engineering. OSS is a human-lead directive combining social coordination with technical innovation, and as a socio-technical field is fit for application of these techniques to capture inherent structure such as identified by \textit{Engagement} in \autoref{fig:engagement}.

EFA assumes that the latent construct (e.g. \textit{Robustness}) is responsible for the correlation of the indicator variables. In practice this allows the researcher to conclude on a statement of influence, i.e.: “developer engagement is positively related to pull requests.” This method does not assume perfect measurement of observed variables and allows the factor to explain what the indicators have in common, and what is not held in common is due to measurement error~\cite{Clark2018}.

The EFA approach is used here because our goal is identify latent constructs of \textit{Interest} and \textit{Robustness} for building a theory of OSS blockchain health~\cite{Fabrigar2012}. Additionally, EFA allows for correlation between latent variables which is to be expected in the confirmatory factor analysis stage of structural equation modelling (next). We follow the approach of Hair~\cite{Hair2014} for EFA by following the six stage framework, and continue with the six stages for SEM.

\subsection{Structural Equation Modelling}\label{sec:SEM}
Exploratory factor analysis can be used to find underlying structure in a collection of variables that possibly represent \textit{Interest} and \textit{Robustness}. To assess the nature of the relationship between these two latent factors, structural equation modelling (SEM) can be applied.  SEM is employed both for the development of the measurement model and the evaluation of its structural efficacy. The structure of the relationship emerges similar to a regression, and determines the strength of the relationships between constructs, although with SEM there is no predetermined directionality. Taking empirical data--from GitHub--and testing it against a theoretical model is the key benefit of this technique~\cite{ChenF2008}.

Although prevalent in sociology, economics, and psychology, very few multivariate statistical approaches such as EFA and SEM have been applied to OSS. Chengalur-Smith et al.~\cite{Chengalur2010} used SEM to model longitudinal project sustainability based on an ecological model; Abdulhassan Alshomali~\cite{Abdulhassan2018} modelled trends in GitHub programming languages via SEM; Raja and Tretter~\cite{Raja2012} developed a regression model for software viability; and Schroer and Hertel~\cite{Schroer2009} used SEM via partial least squares path analytic models to investigate the structure of engagement tasks of Wikipedia volunteers. These studies indicate the applicability of SEM applied to OSS data collection, as well as the opportunity based on the gap in the literature.

\subsection{DataSet}\label{sec:dataset}
Data for public blockchain projects are readily available for collection and analysis from GitHub through the web interface, programmatically through the application programming interface (API), and in raw archival form from the GitHub Archive.

All GitHub data from February 2011 has been archived in JSON Lines format (JSON object on every line) and is available for public download from the GHArchive\footnote{\url{https://www.gharchive.org/}} project. Every JSON object contains the metadata and payload for one GitHub event. For example, when a repository is starred an event is emitted of \texttt{type: stargazer}. Similarly an event is created when a pull request (PR) is created, and the JSON contains all the details about who created it, when it was created, and the contents of the PR object.

GitHub currently has 17 event types\footnote{\url{https://docs.github.com/en/developers/webhooks-and-events/events/github-event-types}}, seven of which are used to collect relevant data: \texttt{WatchEvent}, \texttt{ForkEvent}, \texttt{PushEvent}, \texttt{PullRequestEvent}, \texttt{IssueCommentEvent}, \texttt{CommitCommentEvent}, and\\ \texttt{PullRequestReviewCommentEvent}. All the events have metadata with \texttt{author username}, \texttt{date}, and \texttt{time} which can be used for further metrics.

GHArchive data was downloaded from 01-February-2011 up to 26-March-2022, consisting of 2.36~TiB in total information. This forms the basis of the GHArchive-sourced metrics shown in \autoref{tab:efa-metrics}. The compressed JSON is then inserted into a single-table ClickHouse database~\cite{Milovidov2020}. ClickHouse\footnote{\url{https://clickhouse.com/}} is an open source column-oriented database management system designed for online analytical processing. This is ideal for large datasets that involve mostly read-only queries and batch updating. The database contains 5.6 billion records and is 430~GiB and is accessed with structured query language (SQL) queries through a command line or a Python module. This is running on a dedicated Linux Ubuntu (version 20.04.4) machine with ClickHouse’s command line client and server (versions 22.3.3.44) installed.

To identify relevant blockchain projects the top 600 are gathered using the CoinMarketCap API by ranking of market capitalisation as of March-2022. Details retrieved include project \textit{name}, \textit{rank}, \textit{website}, and \textit{location of source code} if available. The data is collated into a Pandas dataframe (version 1.4.2) for Python (version 3.8.10) via JupyterLab (version 3.3.3). The CMC data provides a rank based on total market of a given project which can be a proxy for financial resources. The website information is then used with Amazon's Alexa API to get a global web ranking called Alexa Traffic Rank\footnote{Run by Amazon's subsidiary Alexa Internet, Inc., the service was shuttered on May 01, 2022.}.

\begin{table}[h]
	\centering
	\begin{threeparttable}
		\caption{Metrics used for the exploratory factor analysis (\cref{sec:EFA}), brief description of the operationalization and the data source.}
		\label{tab:efa-metrics}{\small
		\begin{tabularx}{\linewidth}{@{}p{3.3cm}>{\hangpara{3mm}{1}}Xp{1.7cm}@{}}
			\toprule
			Metric & Operationalization & Source      \\
			\midrule
			Stars	& Total count of stars since the project's inception & GHArchive\tnote{$\star$}\\
			Forks	& Total count of forks since the project's inception& GHArchive\\
			Alexa rank	& Amazon's Alexa global web rank based on the project's website & Alexa\\
			CMC rank	& CoinMarketCap's rank based on the project's market capitalisation & CMC\tnote{$\ast$}\\
			Mentions	& Total count of project mentions in the commit history & OSSF\tnote{$\dagger$}\\
			Geographic distribution	& Activity based on timezone distribution & custom\\
			Criticality score & Score based on the project's influence and importance&OSSF	\\
			Longevity	& Average number of days the developers have been involved & GHArchive\\
			Last updated	& Number of months since the project has been updated & GHArchive\\
			Median response time & Median number of days for issues to be closed & GHArchive \\
			Average response time& Average number of days for issues to be closed & GHArchive \\
			\bottomrule
		\end{tabularx}}
		\begin{tablenotes}[flushleft]\footnotesize 
			\item[$\star$] GHArchive archives all GitHub data.
			\item[$\ast$] CMC is CoinMarketCap.
			\item[$\dagger$] OSSF is the Open Source Software Foundation.
		\end{tablenotes}
	\end{threeparttable}
\end{table}

Two further metrics are pulled from the Open Source Software Foundation (OSSF): project criticality score, and the number of mentions. The criticality score is a metric that identifies how critical a project is within the open source ecosystem~\cite{OSSF2022}. Scored in the range $[0,1]$ a project of score 0 relies on no external software, among other factors, while a project of score 1 is deemed critically important. For example, the highest criticality project overall is Linux while the highest criticality blockchain project is Bitcoin. Mentions is a metric to gauge what projects are popular among contributors and is based on a count of the number of times a project appears in the text through comments of the commit messages. A Python script is written to access GitHub data through the API via the Criticality Score\footnote{\url{https://github.com/ossf/criticality_score}} command line tool (version 1.0.7).

Geographic distribution was introduced as a measure of robustness from \autoref{tab:metrics-robustness}. This is derived from timezone data retrieved from git history using Perceval~\cite{Duenas2018} (version 0.17.0) via a Python script. The timezone data represents the times of software commits made by the project's contributors and produces a mapping of activity based on coordinated universal time (UTC). To evaluate a project's geographic distribution, we compare it to a median distribution of the top 100 blockchain projects over the previous six months. This median distribution is a representation of the typical geographic distribution of the top 100 blockchain projects in terms of software commit activity. The comparison is done by calculating the root mean squared error (RMSE), which is a measure of the difference between the project's geographic distribution and the median distribution. Projects with a low RMSE have a distribution that matches the community and are less prone to geographic shocks, which refer to unexpected events that could disrupt the project's contributors' ability to work together. Projects with a high RMSE likely indicates a project operates in a single timezone and could exhibit single point of failure risks, which refer to the risk that the project's development could be disrupted if a key contributor is unable to work.

\subsubsection{Data Processing}\label{sec:processing}
In many cases the source code location is incorrectly reported and so all project source code locations are manually verified. Where the location points to an organization on GitHub, the repository (repo) with the \textit{reference},  or \textit{core}, or \textit{node} implementation is chosen. If this is not the case (perhaps it is not a blockchain), then the \textit{contract} repository is chosen. To disambiguate between competing repos, the one with the most stars is chosen. Often the core repo is also the one with the most stars. When there are two implementations in different languages (e.g., GO and Rust) highest stars takes preference. Forked libraries are not considered, if the original library is in the top 600, it is counted.

As an example of verification, the Sythetix ecosystem has six pinned repositories and is listed by CoinMarketCap as having code at \texttt{\url{https://github.com/Synthetixio}} however this is the organization landing page and contains links to all repos. The main platform is hosted at \texttt{\url{https://github.com/Synthetixio/synthetix}}, which is manually verified.

The top 600 blockchain projects are the starting point and the dataset is cleaned with the following exclusions:
\begin{enumerate}
	\item The version control data cannot be accessed for analysis. Eight have a repo that’s missing (404 error) indicating it has been deleted or moved; 78 are listed but private; 83 do not have a repo listed (and are likely private).
	\item Four are hosted on GitLab\footnote{\url{https://gitlab.com/}}, and two on Bitbucket\footnote{\url{https://bitbucket.org/}}. These GitLab and Bitbucket ones are excluded because they are a small percentage of the whole (1.4\%) and would require separate infrastructure to access the code bases.
	\item Six are doubles where the project points to the same code base for a related project, e.g. KavaSwap (SWP) and Kava (KAVA); only the highest ranked project is included.
\end{enumerate}

Of the 419 publicly available repos on GitHub, 26 of these have no contribution history indicating the repo was created or the code was copied there and never updated. These are excluded as they can be considered dead by ecological standards or stagnant by software measures.

The dataset under investigation contains missing data for 15 projects, spanning three categories: last updated, mentions, and criticality score. Given that these missing values represent 3.9\% of the entire sample ($15/384$), implementing a data imputation strategy can be justified. Two additional projects have a missing Alexa Rank (0.5\%). The chosen method for this task is mean substitution, a common and widely recognized technique for dealing with missing data~\cite{Hair2014}.

The data has been examined for outliers and although there are extreme values, these observations represent actual blockchain projects and are included in the spirit of producing a representative model. Exclusion of potential outliers is tricky, and without good cause such as impossible values or missing data it is best to include them~\cite{Aguinis2013}.

The final dataset contains full data for 384 blockchain projects on which this analysis is based. The descriptive statistics are shown in \autoref{tab:desc-stats}.

\begin{table}[h]
	\centering
	\begin{threeparttable}
		\caption{Descriptive statistics for the dataset used in exploratory factor analysis to identify latent constructs of \textit{Interest} and \textit{Robustness}.}\label{tab:desc-stats}{\small
	\begin{tabular}{@{}l
			S[table-format=6.1]
			S[table-format=6.1]
			S[table-format=3.2]
			S[table-format=5.1]
			S[table-format=6.1]
			S[table-format=6.1]
			S[table-format=7.1]
		}
		\toprule
		$n$=384 							& {Mean} 	& {Std.Dev.} & {Min.} 	& {Q1} 	& {Median} 	& {Q3} 	& {Max.} \\
		\midrule
		Forks 								& 539.64 	& 3437.34 	& 0.00 	& 12.50 & 60.00 	& 206.50 & 59013 \\
		Stars 								& 790.21 	& 4399.46 	& 0.00 	& 24.00 & 107.00 	& 413.50 & 72112 \\
		Mentions 							& 2509.33 	& 26945.39  & 0.00 	& 0.00  & 15.00 	& 156.00 & 492320 \\
		Criticality 						& 0.35 		& 0.18 		& 0.02 	& 0.20  & 0.37 	& 0.49 	& 0.85 \\
		Last updated\tnote{$\ast$} 			& 6.63 		& 10.66 	& 0.00 	& 0.00  & 2.00 	& 8.00 	& 64 \\
		CMC rank\tnote{$\ast$} 				& 271.15 	& 169.63 	& 1.00 	& 120.50 & 260.50 	& 409.00 	& 600 \\
		Geographic distribution\tnote{$\ast$}		& 0.36		& 0.05 		& 0.15 	& 0.33  & 0.38 	& 0.40 	& 0.47 \\
		Longevity 							& 191.32 	& 131.39 	& 0.00 	& 101.02  & 178.14 	& 256.38 	& 763.80 \\
		Alexa rank\tnote{$\ast$} 			& 296412.46 & 552446.14 & 110	& 43913.50 & 131243.50 & 290192.00 & 4628993 \\
		Median resp. time\tnote{$\ast$} 	& 18.97		& 29.56		& 0.00 	& 0.76	  & 2.54 	& 22.95 	& 75.74 \\
		Average resp. time\tnote{$\ast$} 	& 30.28		& 36.61 	& 0.00 	& 5.87    & 12.43 	& 33.80 	& 100.12 \\
		\bottomrule
	\end{tabular}}
		\begin{tablenotes}[flushleft]\footnotesize
			\item[$\ast$] These metrics are reverse-scored so that the larger number has a positive association.
		\end{tablenotes}
	\end{threeparttable}
\end{table}

\section{Results}\label{sec:results}
Cross referenced with the research questions, the results are now presented beginning with \textbf{RQ1} which asked What facets make up a high-level definition of software health, without using or defining specific metrics? We defined health in \cref{sec:health-definition} and summarized in \autoref{fig:health-tree} as a combination of the latent constructs sustainability, robustness, and niche fit. Sustainability is further divided into the factors of \textit{Engagement} and \textit{Interest}.

The second research questions, \textbf{RQ2}: Given that software \textit{Robustness} is a contributing component of health, what are the factors that contribute to software robustness? and \textbf{RQ3}: How does general popularity, or \textit{Interest} fit in? are answered in the exploratory factor analysis in \cref{sec:results-EFA}.

\textbf{RQ4} investigated the nature of the relationship between the components contributing to a definition of software health in \cref{sec:results-CFA}. Finally, the model validation is in \cref{sec:validation-cfa}.

\subsection{EFA for Interest \& Robustness}\label{sec:results-EFA}

Starting with the data in \autoref{tab:desc-stats} a Scree plot and parallel analysis have been examined to see the number of proposed factors. Both a standard scree and parallel analysis indicate preference for a two factors with eigenvalues $\lambda_1=3.11$ for the first factor and $\lambda_2=1.84$ for the second. A parallel analysis provides more robust reasoning as it will calculate the eigenvalues of the observed data and compare them to the eigenvalues of randomly generated data. Significant deviation means there is grounding for grouping by factors. The simulated groups from PA have $\lambda_1=0.52$, and $\lambda_2=0.23$

Factor analysis was carried out with the Psych package (version 2.2.3) in \textbf{\textsf{R}} (version 4.0.2). The computational method used to estimate the factors is maximum-likelihood, or ML, known to perform well when the factor-variable relationships are strong. The principle-axis method was also used for comparison purposes as it is ideal for non-normality and small sample sizes~\cite{Watkins2018} to no significant difference. Factor rotation is done with the GPArotation package (version 2022.4-1). The \textsc{Varimax} factor rotation method maximizes the variances of the loadings within the factors. This can help with structure for two or more factors.

A first iteration of EFA is carried out and both \textit{average-} and \textit{median-response time} do not load onto Factor 1, having a mild negative influence. They load strongly on an independent factor consisting just of themselves which does not meet the criteria for inclusion as they measure roughly the same thing. They are then excluded as part of the EFA iteration process. This is discussed further in \cref{sec:results-EFA}. The Bayesian information criterion (BIC) is a comparator between models, and the BIC improves significantly from $276.252$ to $-36.737$ with their exclusion from the analysis. \autoref{tab:EFA-loadings} shows the EFA results.

\begin{table}[t]
	\centering
	\caption{Exploratory factor analysis loadings for two latent factors 1 and 2, and common variance, $h^2$. Strong indicator  relationships are in bold. The variables \textit{longevity} and \textit{Alexa rank} do not exhibit enough influence to be included in either latent construct.}
	\label{tab:EFA-loadings}{\small
	\begin{tabular}{	@{}
			l
			S[table-format=1.3]
			S[table-format=1.3]
			S[table-format=1.3]
			@{}}
		\toprule
		Indicator & {Factor 1} & {Factor 2} &{ $h^2$} \\
		\midrule
		Forks               	& \textbf{0.988} & 0.137 & 0.995  \\
		Stars               	& \textbf{0.970} & 0.166 & 0.968  \\
		Mentions            	& \textbf{0.885} & 0.076 & 0.790  \\
		Criticality score   	& 0.135 & \textbf{0.988} & 0.995  \\
		Last updated        	& -0.015 & \textbf{0.705} & 0.498  \\
		CMC rank            	& 0.169 & \textbf{0.373} & 0.167  \\
		Geographic distribution & 0.082 & \textbf{0.369} & 0.143  \\
		Longevity	 	    	& 0.075 & 0.237 & 0.062  \\
		Alexa rank      	    & 0.163 & 0.104 & 0.037 \\
		\midrule
		\textit{SS loadings }          & \textit{2.787} & \textit{1.868} \\
		\textit{Cumulative variance}   & \textit{0.310} & \textit{0.517} \\
		\textit{Proportion explained}  & \textit{0.599} & \textit{0.401} \\
		\bottomrule
	\end{tabular}}
\end{table}

With EFA, all measured variables are related to every factor by a factor loading estimate where $-1$ is a strong negative, 0 is neutral, and $+1$ is strong positive relationship. The significant loads for each factor are shown in boldface along with their loading on the secondary factor. There is strong support for a first factor of \textit{Forks}, \textit{Stars}, and \textit{Mentions}, with a second factor of \textit{Criticality score}, \textit{Last updated}, \textit{CMC rank}, and \textit{Geographic distribution}. The chosen cutoff point for loading estimates is 0.3, and thus \textit{Longevity} is out of range and does not have enough influence to describe either factor. The same applies to \textit{Alexa rank}, both of which are excluded for the measurement model (\cref{sec:results-CFA}).

The communalities ($h^2$ in \cref{tab:EFA-loadings}) or common variance refers to the proportion of variance in the observed variables that is explained by the corresponding latent factor. It represents the proportion of the variance in the observed variable that can be attributed to the factor, after accounting for measurement error. For instance, Stars has an $h^2$ of $0.968$ for the Factor 1, indicating that 96.8\% of the variance in stars data can be explained by the first latent factor.

\textit{SS loadings} represents the amount of variance in the observed variables that is accounted for by each factor. Here, the SS loadings for Factor 1 and Factor 2 are 2.783 and 1.920, respectively, indicating that Factor 1 accounts for 2.787 units of variance in the observed variables, and Factor 2 accounts for 1.868 units of variance. The \textit{Cumulative variance} refers to the total amount of variance in the observed variables that can be explained by the factors up to that point. For two factors in the data over half of variance (0.517) is accounted for in the model with 59.9\% of it coming from Factor 1. The remaining 40.1\% is from Factor 2.

\begin{figure}[h]
	\centering
	\includegraphics[trim={0.3cm 0.3cm 0.3cm 0.3cm}, clip, width=1.0\linewidth]{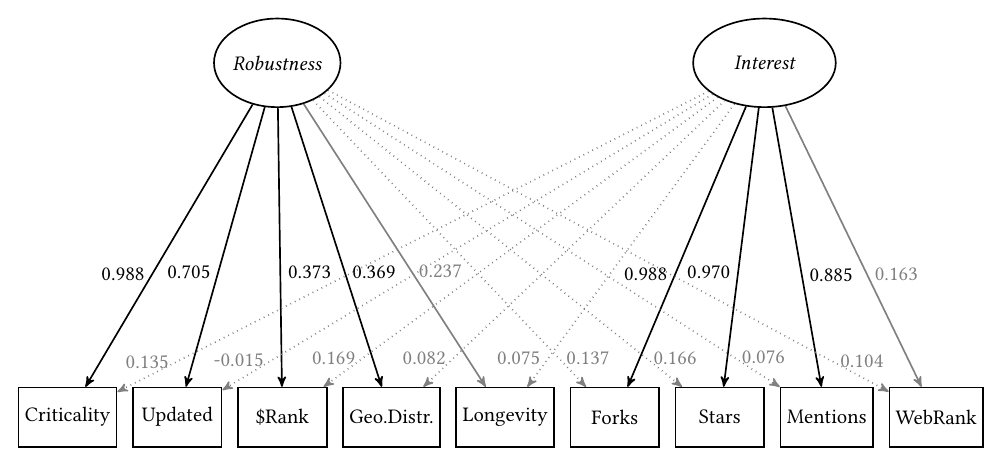}
	\caption{Exploratory factor analysis applied to metrics representing \textit{Robustness} and \textit{Interest}. The primary loadings are the solid arrow in the first tier; the secondary loadings the dotted arrows in the second tier. Note: \textit{median} and \textit{average response time} have already been excluded from the model. \textit{Longevity} and\textit{ Web Rank} are excluded at this stage.}
	\Description{Exploratory factor analysis applied to metrics representing robustness and interest.}
	\label{fig:efa}
\end{figure}

\autoref{fig:efa} shows the EFA diagram with strong loadings in black and secondary loadings in grey. This model exhibits fit statistics in the range of standard thresholds: the Tucker Lewis Index of factoring reliability (TLI) $=0.953$, and the root mean square error of approximation (RMSEA) index $=0.089$. See \cref{tab:valid} for these in context of the model validation. More on fit statistics is discussed in \cref{sec:discussion}.

\subsubsection{EFA Validation}\label{sec:validation-efa}
Model validation is by two mechanisms. First cross-validation of the EFA by randomly separating the dataset into a training and testing segment and comparing model structure. Secondly, confirmatory factor analysis is applied to the measurement model (\autoref{sec:SEM}).

Cross validation is necessary to avoid the situation where the model ends up being overfit to the data, affecting generalizability. Constraints in the data collection process on the number of available projects limit collecting an entire new dataset and so the original is split into two groups. Random allocation is performed using the Caret package (version 6.0.86) with 51\% split to produce two groups--one to build the model and one to test the model. This split allows for half the data to be at the minimum sample size threshold of 200.

Exploratory factor analysis is used to define the latent constructs of \textit{Interest} and \textit{Robustness}. \autoref{tab:valid} shows the validation results. The insignificant loadings are not shown for readability and to highlight that the same factor structure is present across the models. The training and testing models are both as well fit as the hypothesized Model based on $\chi^2$, TLI, and RMSEA, with slight deviations being acceptably close considering the sample size limitation.

\begin{table}[h]
	\centering
	\caption{Cross-validation for the Exploratory Factor Analysis model showing equivalent factor structure for the testing model as compared to the baseline EFA. Model grouping 1 is \textit{Interest} and grouping 2 is \textit{Robustness}. Insignificant loadings ($<0.3$) are not shown except where appropriate for structure comparison.}
	\label{tab:valid}{\small
	\begin{tabular}{llllllllll}
		\toprule
		&   & \multicolumn{2}{c}{Model } &   & \multicolumn{2}{c}{Training} &   & \multicolumn{2}{c}{Testing}  \\
		\cmidrule{3-4}\cmidrule{6-7}\cmidrule{9-10}
		Factor        & \phantom{a} & 1   & 2                   & \phantom{a} & Tr1   & Tr2    & \phantom{a} & Te1   & Te2  \\
		\cmidrule{1-1}\cmidrule{3-4}\cmidrule{6-7}\cmidrule{9-10}
		Forks       	&  	& 0.988 &        &   & 0.983 &      &    & 0.964 &            \\
		Stars      		&  	& 0.970 &        &   & 0.976 &      &    & 0.962 &            \\
		Mentions 		&   & 0.885 &        &   & 0.924 &      &    & 0.910  &            \\
		Criticality     &   &       & 0.988  &   & 		 & 0.994 &   &     	  &  0.980     \\
		Last updated 	&   &       & 0.705  &   &       & 0.707 &   &        &  0.711     \\
		CMC rank 		&  	&      	& 0.373  &   &       & 0.425 &   &    	  &  0.292     \\
		Geographic distribution & & & 0.369	&   &        & 0.398 &   &    	  &  0.342     \\
		Longevity	 	&   &		& 0.237  &   &       & 0.288 &   &    	  &  0.197      \\
		Alexa rank      &   & 0.163	&       &   &  0.07 &  		 &   & 0.419   &           \\

		\cmidrule{1-1}\cmidrule{3-4}\cmidrule{6-7}\cmidrule{9-10}
		$n$           &   & \multicolumn{2}{c}{388}     &   & \multicolumn{2}{c}{213}      &   & \multicolumn{2}{c}{171}     \\
		$\chi^2$      &   & \multicolumn{2}{c}{2373.7}  &   & \multicolumn{2}{c}{1680.1}  &   & \multicolumn{2}{c}{938.8} \\
		TLI           &   & \multicolumn{2}{c}{0.953}   &   & \multicolumn{2}{c}{0.972}    &   & \multicolumn{2}{c}{0.969}   \\
		RMSEA         &   & \multicolumn{2}{c}{0.089}   &   & \multicolumn{2}{c}{0.077}    &   & \multicolumn{2}{c}{0.067}   \\
		\bottomrule
	\end{tabular}}
\end{table}

At this stage we can officially rename the latent constructs to be representative of the underlying indicator variables, thus Factor 1 becomes \textit{Interest}, and Factor 2 becomes \textit{Robustness}.

\subsection{CFA Model}\label{sec:results-CFA}
We now have latent variables representing \textit{Engagement}, \textit{Interest}, and \textit{Robustness}. To investigate the nature of the relationship between the components (\textbf{RQ4}), a measurement model is first hypothesized on the basis of the previous EFA and validated using confirmatory factor analysis (CFA).

\begin{figure}[h]
	\centering
	\includegraphics[trim={0.3cm 0.3cm 0.3cm 0.3cm}, clip, width=1.0\linewidth]{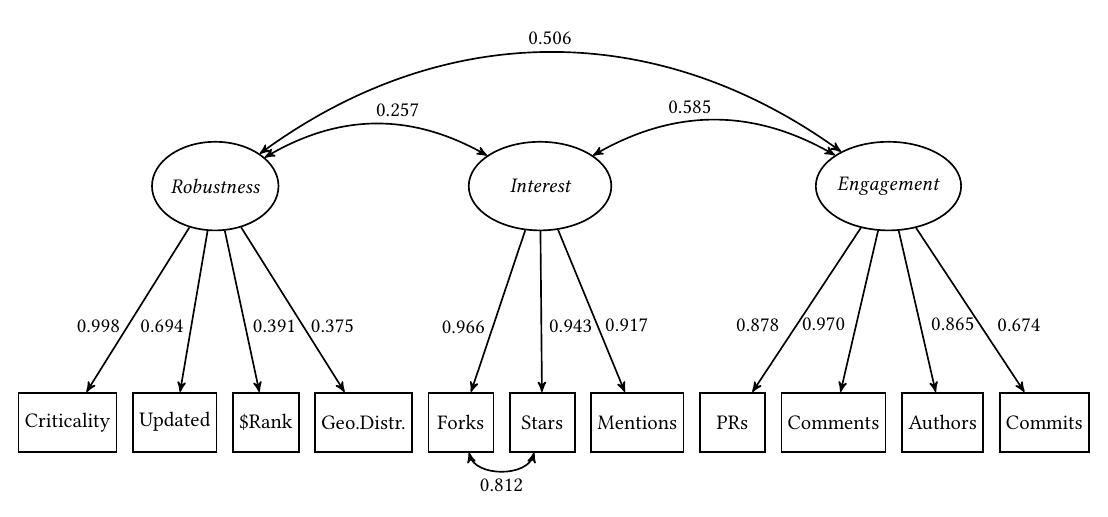}
	\caption{Measurement model baseline showing correlations between latent constructs and forks \& stars. Loadings are standardized across all eleven indicator variables.}
	\Description{Confirmatory factor analysis applied to latent factors of interest, robustness, and engagement.}
	\label{fig:cfa}
\end{figure}

The EFA model from \autoref{fig:efa} is combined with the Engagement model (\autoref{fig:engagement}) to produce the  measurement model seen in \autoref{fig:cfa}. The CFA and SEM analysis was carried out with the Lavaan package (version 0.6.13) in \textbf{\textsf{R}} (version 4.0.2). During the EFA \textit{response time} was eliminated, and at this stage two more indicators are removed: \textit{longevity} and \textit{web rank} because they fell below our 0.3 threshold. There are no cross-loadings in the CFA since it limits the loading to the theoretical construct only, setting any remaining loadings to zero.

A correlation path between Forks and Stars has been freed to improve the model by eliminating a potential Heywood case. A Heywood case is a factor that has a negative error variance estimate and can result in an improper computational solution~\cite{Hair2014}. Less than zero error is illogical as it means that more than 100\% of the variance in the data is due to the factor structure. The alternate remedy here is to remove Forks altogether as a metric, however, we chose to keep the variable because first it allows for three indicators on \textit{Interest} (dropping to two indicators is considered under-fit), and second it is an important concept in OSS.

\subsubsection{Measurement Model Validation}\label{sec:validation-cfa}
The second method to validate the model is by confirmatory factor analysis on the measurement model (\autoref{fig:cfa}). Here the data shows good support for the hypothesized model indicating the latent constructs are represented by their indicator variables. However, the goodness of fit statistics indicate there is room for model improvement. The Comparative Fit Index (CFI) is 0.84 and the Tucker Lewis Index (TLI) is 0.78 both of which are under a 0.90 heuristic threshold, and the RMSEA is 0.214 $(>0.07)$ and SRMR is 0.106 $(>0.08)$, both of which are over heuristic thresholds meaning there is evidence for good model fit, but not great. These fit statistics must be used cautiously as there is evidence that with sample sizes approaching 400 the maximum likelihood estimator becomes sensitive to changes in the data resulting in poor fit~\cite{Hair2014}. This is to be explored further in \cref{sec:limitations}--\nameref{sec:limitations}.

Internal consistency reliability is measured to ensure that the items in our scale are measuring the same construct consistently which increases our confidence in the validity of the scale. Estimates of internal consistency reliability for each scale based on Cronbach's alpha~\cite{Cronbach1951} and McDonald's omega~\cite{Mcdonald1999} coefficient shows high levels of internal consistency reliability for all scales with alpha coefficients ranging from 0.69 to 0.97, indicating good to excellent reliability. The omega coefficients are also high, ranging from 0.73 to 0.95, indicating good to excellent general factor saturation. These estimates suggest that the scales are reliable measures of the constructs they are intended to measure. in other words the data from the selected metrics are accurately captured in the latent variable.

\subsection{Proposed Structural Model}\label{sec:results-SEM}
\begin{figure}[h]
	\centering
	\begin{subfigure}{0.43\textwidth}
		\centering
		\includegraphics[width=\textwidth]{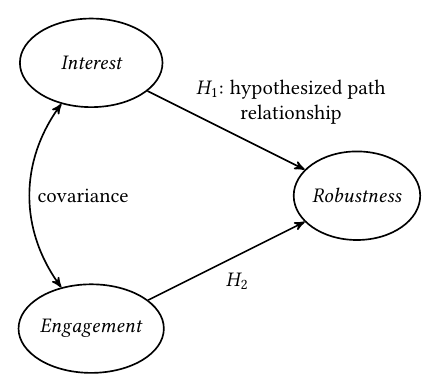}
		\caption{Proposed structural model with hypothesized path relationships.}
		\label{fig:sem-0-subfig}
	\end{subfigure}
	\hspace{0.7cm}
	\begin{subfigure}{0.43\textwidth}
		\centering
		\includegraphics[width=\textwidth]{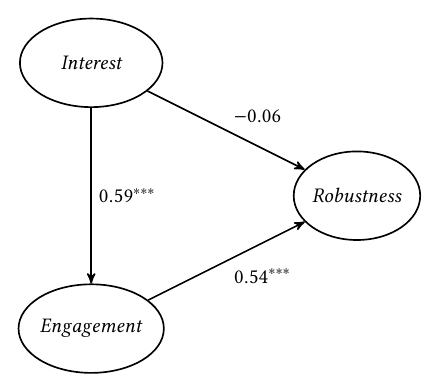}
		\caption{Resulting structural model with path estimates.}
		\label{fig:sem-1-subfig}
	\end{subfigure}
	\caption{Structural model showing insignificant effect of \textit{Interest} on \textit{Robustness}. \textit{Interest} is a contributing factor to \textit{Engagement}, which, in turn is a contributing factor to \textit{Robustness}. Indicators and loadings are not shown and are consistent with \autoref{fig:cfa}. $^{***}$ is statistically significant at $p<0.001$.}
\Description{Proposed structural equation model.}
\label{fig:sem}
\end{figure}

The proposed structural model is derived from the definition of software health (\cref{sec:health-definition}) as composed of our three latent constructs found in the EFA. \autoref{fig:sem-0-subfig} shows a structural model where general \textit{Interest} leads to software \textit{Robustness} and an independent path of \textit{Engagement} $\rightarrow$ \textit{Robustness}. These paths are the hypothesized relationships. $H_1$ says that an increase in Interest will positively inform Robustness. $H_2$ is that there is a positive influence of Engagement on Robustness.

SEM results show a high correlation between \textit{Interest} and \textit{Engagement} suggesting there might be an underlying structural relationship between them. As interest represents the lighter touch activities such as starring a repository, it is reasonable to suggest that this activity predicts engagement or more developer-centric activities such as committing to a code repository. The structural relationship is seen in the revised model in \autoref{fig:sem-1-subfig}.

The SEM results in \autoref{fig:sem-1-subfig} show there is almost no effect of \textit{Interest} on \textit{Robustness} (loading of $-0.06, p>0.05$) and thus this relationship, $H_1$, is not supported. The second hypothesis, $H_2$, has strong support that \textit{Engagement} predicts \textit{Robustness} (loading of $0.54, p<0.001$). The third result here is that general Interest is a leading indicator of developer Engagement (loading of $0.59, p<0.001$). The model was revised to remove the path relationship $H_1$ and results were similar showing \textit{Interest} $\rightarrow$ \textit{Engagement} strength 0.58, and \textit{Engagement} $\rightarrow$ \textit{Robustness} strength 0.50.

\subsubsection{Structural Model Validation}\label{sec:validation-sem}
The structural model shown in \autoref{fig:sem-1-subfig} specifies the path relationship between the constructs. At this point the nomological validity is evaluated to see if the test measures what it should be measuring. We have already used the literature to inform the factors, and so this test is a conceptual check on the outcome relationships. After removing the path between \textit{Interest} and \textit{Robustness}, there are two remaining paths:
\begin{enumerate}

	\item \textit{Interest} $\rightarrow$ \textit{Engagement} $\,$ The interest construct is made up of Forks, Stars, and Mentions which are closer to social metrics than traditional software development. For example, someone that is interested in Bitcoin is likely to first star the repository as a bookmarking method, then fork it if they are further curious. These activities happen before any discussion about bugs or changes, and before any new code is written and a pull request submitted. So, in the procedural sense of contributing to OSS, interest leads, or predicts, engagement.
\medskip
	\item \textit{Engagement} $\rightarrow$ \textit{Robustness} $\,$ Going back to our ecology metaphor, an organism must be able to sustain its base metabolic needs for survival before it can grow and thrive in its environment. Only once the basic needs are met can it become strong enough to survive shocks and adapt to changes in the environment. In software, this base survival is the day-to-day operations and involves communicating with community members, writing code, submitting code reviews, and attending to comments. If these needs are met through the contributors' motivation and satisfied working conditions, etc., then the project can be in a position to strengthen against unknown future disruptions.

\end{enumerate}

\section{Discussion}\label{sec:discussion}

\subsubsection*{\textbf{Niche Fit Metrics}}
From the baseline definition of software health that has a parallel in ecosystem health (\cref{sec:health}) the local context of a species or project in the ecosystem is deemed important. This niche occupancy has sound logic: if a software project fills a specific market gap and has no competition it is positioned to thrive, and, more likely to be healthy. The metrics in \autoref{tab:metrics-niche} present the researcher with the complex issue of how exactly to identify the niche and quantify it. The perspective taken here is that of the individual software project which limits the context required to determine if it fits a niche (is unique) or not (has strong competitive alternatives, etc.). In other words, we are not analysing the local environment to see if what the project delivers fills a niche, rather the empirical approach is content agnostic, and seeks to determine health without the subjective approach of determining market fit, or other such niche indicators. As such, there is limited research to operationalize niche metrics in OSS.

Chengalur-Smith et al.~\cite{Chengalur2010} defined the construct of niche through audience niche, programming language niche, and operating system niche. What audience niche means is unclear, however language and OS niche are if the project has support for less popular languages and platforms. The study found that none of these metrics had a significant effect on attraction or sustainability, with niche size path estimates of less than 0.04. While this suggests that more research is needed to determine the relationship between niche occupation and software health, it also highlights the complexity of measuring and interpreting these metrics in the context of collaborative software engineering.

This presents as a limitation to the current study, as we have no indicators for niche, our overall model for health might be under-identified, and possibly contribute to the weak model fit indices in \cref{sec:validation-cfa}.

\subsubsection*{\textbf{Where does general interest or popularity fit into software health?}}
Jansen~\cite{Jansen2014} categorizes interest as part of robustness whereas we have found interest to be a part of sustainability and found it has no direct influence on robustness. The \textit{Interest} $\rightarrow$ \textit{Robustness} loading is insignificant at $-0.06$ (\cref{sec:results-SEM}).

This raises questions about the impact of popularity metrics on software health. While it seems interest can contribute to the robustness of a project by increasing its popularity and attracting more developers, it is more likely that increased popularity increases engagement which then affects robustness. One of the benefits of structural equation modelling is being able to disambiguate this relationship.

Forks and stars are strong interest metrics, with both being used by Osman and Baysal~\cite{Osman2021} to define \textit{popularity}, and by Abdulhassan~\cite{Abdulhassan2018} to define \textit{repo interest}. Additionally, Negoita et al.~\cite{Negoita2019} have stars as a sole definition of \textit{sustainability}. As Jansen points out, once a competitor emerges, users may shift their attention to a more promising alternative, potentially causing long-term damage to the original project. In this sense, the concept of a popular project may align more closely with sustainability, rather than robustness.

\subsubsection*{\textbf{What happened to bug-fix time?}}
Based on the literature, bug fix rate is one of the most cited measures in software health (\cref{tab:metrics-sustainability}; as a part of sustainability). In the present study there is no evidence within this blockchain dataset that time to fix bugs, as an independent measure, is crucial. The original exploratory factor analysis data includes two metrics to gauge their influence: the median time to fix bugs based on the time delta between issues being opened and closed, as well as the average value. Both median and average were investigated as it was thought median value would be more beneficial to account for the long tail in issues that are not critical waiting until someone has the time to explore them. That model did not incorporate median and average time into a structural relationship, rather those two indicators stood alone in an independent factor. As they approximately measure the same concept (improving software by responding to issues) the indicators should not both be included as an under-identified factor. Thus all direct measurement of issue close rate is absent. One explanation is that attending to issues is one of the prime activities that contributors work on, especially new members that looking for a place to start contributing~\cite{Wang2020}. To get started in a new community they can easily browse the list of open issues to see what needs to be done, thus engaging with the project by fixing a bug. Its not that this activity is not important, rather it is already captured within engagement through commits, comments, and pull requests. From the researcher's point of view this yields a more parsimonious model by reducing indicator redundancy.

\subsubsection*{\textbf{Is healthy software robust software?}}
Goggins et al.~\cite{Goggins2021} agree with Chengalur-Smith et al.~\cite{Chengalur2010} in their definitions of health as a combination of \textit{sustainability} and \textit{survivability} as shown in \cref{tab:health}.

Our concept mapping chose to use the term robustness rather than survivability, and if we use these as synonyms for a moment we can see the path structure. A project can be sustainable but not survive. However, a project cannot survive without being sustainable. Surviving projects must therefore also be sustainable. This is supported by our path relationship \textit{Engagement} (part of sustainability) $\rightarrow$ \textit{Robustness} i.e. survivability (\autoref{fig:sem-1-subfig}).

\begin{figure}[h]
	\centering
	\includegraphics[trim={0.3cm 0.3cm 0.3cm 0.3cm}, clip, width=0.7\linewidth]{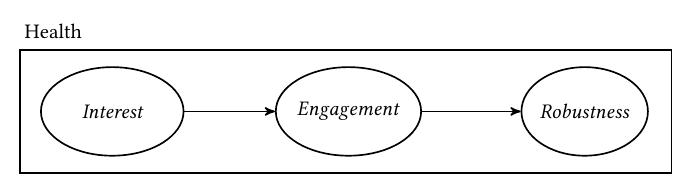}
	\caption{Healthy software is a combination of latent factors, including \textit{Interest}, \textit{Engagement}, and \textit{Robustness}.}
	\Description{Model of Healthy software as a combination of latent factors, including \textit{Interest}, \textit{Engagement}, and \textit{Robustness}.}
	\label{fig:health-model}
\end{figure}

Health as a latent construct is not part of the model as there are no direct indicator metrics that asses health, rather as we have shown the latent variables of interest, engagement, and robustness. So taken together, these represent a picture of health. It is reasonable to assume that robust software is also healthy software, but there is more to it; robustness itself is composed of indicator metrics and an endogenous construct. A more accurate picture of health is in \autoref{fig:health-model}.

\subsection{Implications}\label{sec:}
It is our goal here to solidify some of the research into OSS health in a manner that can provide a clear definition of, and metrics to asses, software health. As Goggins et al.~\cite{Goggins2021} say, ``There is a considerable amount of research constructing and presenting indicators of open source project activity, but a lack of consensus about how indicators derived from trace-data might be used to represent a coherent view of open source project health and sustainability.''

The study results allow stakeholders such as future and current OSS contributors, researchers, and project managers to identify areas for improvement in their software projects. By understanding the factors that contribute to software health, project managers can make informed decisions about where to allocate resources to improve software based on operationalized metrics attributed to interest, engagement, and robustness. The study provides a definition of software health and an introductory structural equation model in the field of blockchain software health. This model can be used as a starting point for future research in this area and can help to guide the development of more comprehensive models of software health. A clear next step will be extension beyond blockchain OSS (\cref{sec:future-work}).

With regards to research into blockchain software, there are a few points to present. The nature of open source contributions allows developers to self-select projects that have an ideological fit~\cite{Smirnova2022}. Perhaps this has implications for the wider software industry, as it suggests that developers are more likely to contribute to projects that align with their personal beliefs and values. In the case of blockchain-based projects, for example, research has shown that developers are more likely to cite motives for contributing based on a "bitcoin ideology" than developers in non-blockchain domains~\cite{Bosu2019, Hars2001}. This suggests that the blockchain industry may attract developers with a particular set of values and beliefs, which could guide newcomers looking to contribute to blockchain open source software.

In addition to ideological factors, blockchain-based projects often rely on token incentives to motivate and reward developers. The suggestion here is that incentive-based participation may be more effective than purely voluntary contributions. While most OSS projects are built on voluntary contributions, grants, and scholarships, blockchain-based projects have the additional incentive of compensation directly through or indirectly through the token economy. This creates a link between the quality of a developer's contribution and a potential financial reward, which may encourage more developers to contribute and improve the overall quality of the project.

\subsection{Limitations}\label{sec:limitations}
Beginning with data collection, a few limitations are notable. First, the data is only sourced from GitHub. This is due to the prominence of blockchain projects being hosted here, but must be considered as there are alternatives. Secondly, repository owners could have moved and renamed repositories in the time between manual verification of the code location and the database query time. Although we have no known instances of this, there is the possibility, for example, that individual projects recorded no or little activity because they were moved from the tagged location.

The study sample size ($n=384$) was acceptable, but for cross validation of the exploratory factor analysis (\cref{sec:validation-efa}) the testing/training split was under 200 as recommended by Hair et al.~\cite{Hair2014}. Increasing the sample size may be beneficial, but it's important to note that simply adding more blockchain projects does not yield an increase in active projects, as old repositories will remain stagnant and their data available long after the project has been deemed dead. Another factor to consider when increasing the sample size is the detrimental effects on the fit indices as the maximum likelihood estimator is sensitive  to variations in the dataset resulting in poor fit.

For statistical validation the fit indices of the confirmatory factor analysis discussed in \cref{sec:validation-cfa} are, as a rule of thumb, used to proceed to the structural equation model that estimates the path relationship coefficients. There is substantial literature in the social sciences regarding fit statistics~\cite{ChenF2008,ClarkD2018,Watkins2018}, and it is best to keep in mind that individual fit statistics are not hard rules, nor provide enough evidence to be used in isolation and therefore it is recommended to use a suite of tests for comparison. Also of note considering the context of software engineering, the recommended thresholds must not be considered law since they were developed based on normally distributed data~\cite{Finch2020} that was usually collected from surveys designed by researchers. Acceptance or rejection of models should not be based on fit statistics, rather on the ability of the model to provide structure to the data.

Lastly, in the health literature there is varied consensus regarding whether or not the category of niche occupation is a component of health. There is clear agreement of its importance in natural ecosystems, and this is mirrored in business ecosystems. When transitioning into software, not as many studies include niche fit, and of the open source ones in \cref{tab:health}, only two of seven include a related concept. So without much prior work or benchmarking in this area it is difficult to conclude if the absence of niche occupation yields an accurate representation. This highlights one area for future work in software health. Other future directions are mentioned presently.

\subsection{Future Work}\label{sec:future-work}
A good structural model (and measurement model) allows for generalizability, which can be tested across different OSS industries, such as mobile, web, tools, finance, and others. By testing the applicability of a structural model across multiple industry domains, researchers can assess the robustness and generalizability of the model, as well as identify any industry-specific factors that may impact software health. This can help to ensure that the model is widely applicable and can provide useful insights for practitioners across a range of collaborative software engineering contexts.

Beyond validating a sound structural model across industries, the direct application of software projects assessed against the model as a predictor of health is a long term goal. By using SEM to model the relationships between these endogenous concepts of \textit{Interest}, \textit{Engagement}, and \textit{Robustness}, it is possible to make predictions about health based on the model. This can be particularly useful in the software development process, as it allows developers to identify which factors are most important for achieving desirable outcomes and to adjust their processes accordingly, perhaps even identifying successful projects.

\section{Conclusion}\label{sec:conclusion}

In this paper we aimed to investigate the definition of software health in the context of open source software (OSS) and define operational measures that can be used to determine health. Our investigation identifies health as a three-pronged concept comprising sustainability, robustness, and niche occupation. Sustainability is further made up of general interest and engagement. We applied exploratory factor analysis to a dataset to find latent constructs for \textit{Interest} and \textit{Robustness}, which extends the previous work of the authors on \textit{Engagement}. The latent factor of \textit{Robustness} is composed of software criticality score, time since it was updated, market capitalisation ranking, and geographic distribution, while \textit{Interest} is made up of forks, stars, and project mentions. A measurement model was created and validated using confirmatory factor analysis. We proposed a structural model that suggests \textit{Interest} informs \textit{Engagement} (positively), which in turn informs \textit{Robustness} (positively) and estimated the path coefficients. While there is good support for the EFA, further work is required to improve the model fit of the proposed structural model. Overall, this research provides insights into the intricacies of OSS health and lays the foundation for future research in this area.


\begin{acks}
We would like to thank software developers for continuing to create open source software.
\end{acks}
\bibliographystyle{ACM-Reference-Format}
\bibliography{acm_refs}


\begin{thebibliography}{63}


\ifx \showCODEN    \undefined \def \showCODEN     #1{\unskip}     \fi
\ifx \showDOI      \undefined \def \showDOI       #1{#1}\fi
\ifx \showISBNx    \undefined \def \showISBNx     #1{\unskip}     \fi
\ifx \showISBNxiii \undefined \def \showISBNxiii  #1{\unskip}     \fi
\ifx \showISSN     \undefined \def \showISSN      #1{\unskip}     \fi
\ifx \showLCCN     \undefined \def \showLCCN      #1{\unskip}     \fi
\ifx \shownote     \undefined \def \shownote      #1{#1}          \fi
\ifx \showarticletitle \undefined \def \showarticletitle #1{#1}   \fi
\ifx \showURL      \undefined \def \showURL       {\relax}        \fi
\providecommand\bibfield[2]{#2}
\providecommand\bibinfo[2]{#2}
\providecommand\natexlab[1]{#1}
\providecommand\showeprint[2][]{arXiv:#2}

\bibitem[{Abdulhassan Alshomali}(2018)]%
        {Abdulhassan2018}
\bibfield{author}{\bibinfo{person}{Mohammad~Azeez {Abdulhassan Alshomali}}.}
  \bibinfo{year}{2018}\natexlab{}.
\newblock \emph{\bibinfo{title}{{Open source software GitHub ecosystem: a SEM
  approach}}}.
\newblock \bibinfo{thesistype}{Ph.\,D. Dissertation}. \bibinfo{school}{James
  Cook University}.
\newblock
\urldef\tempurl%
\url{https://doi.org/10.25903/5c3eb27776753}
\showDOI{\tempurl}


\bibitem[Aguinis et~al\mbox{.}(2013)]%
        {Aguinis2013}
\bibfield{author}{\bibinfo{person}{Herman Aguinis}, \bibinfo{person}{Ryan~K.
  Gottfredson}, {and} \bibinfo{person}{Harry Joo}.}
  \bibinfo{year}{2013}\natexlab{}.
\newblock \showarticletitle{Best-Practice Recommendations for Defining,
  Identifying, and Handling Outliers}.
\newblock \bibinfo{journal}{\emph{Organizational Research Methods}}
  \bibinfo{volume}{16}, \bibinfo{number}{2} (\bibinfo{year}{2013}),
  \bibinfo{pages}{270--301}.
\newblock
\urldef\tempurl%
\url{https://doi.org/10.1177/1094428112470848}
\showDOI{\tempurl}


\bibitem[Arantes and Freire(2011)]%
        {Arantes2011}
\bibfield{author}{\bibinfo{person}{Flávia~Linhalis Arantes} {and}
  \bibinfo{person}{Fernanda Maria~Pereira Freire}.}
  \bibinfo{year}{2011}\natexlab{}.
\newblock \showarticletitle{Aspects of an Open Source Software Sustainable Life
  Cycle}. In \bibinfo{booktitle}{\emph{Open Source Systems: Grounding Research
  - 7th IFIP WG 2.13 International Conference}}. \bibinfo{pages}{325--329}.
\newblock
\showISBNx{9783642244179}
\urldef\tempurl%
\url{https://doi.org/10.1007/978-3-642-24418-6_26}
\showDOI{\tempurl}


\bibitem[Arya et~al\mbox{.}(2022)]%
        {OSSF2022}
\bibfield{author}{\bibinfo{person}{Abhishek Arya}, \bibinfo{person}{Caleb
  Brown}, {and} \bibinfo{person}{Rob Pike}.} \bibinfo{year}{2022}\natexlab{}.
\newblock \bibinfo{booktitle}{\emph{Open Source Project Criticality Score}}.
\newblock San Francisco, California.
\newblock
\urldef\tempurl%
\url{https://github.com/ossf/criticality_score}
\showURL{%
\tempurl}


\bibitem[Bosu et~al\mbox{.}(2019)]%
        {Bosu2019}
\bibfield{author}{\bibinfo{person}{Amiangshu Bosu}, \bibinfo{person}{Anindya
  Iqbal}, \bibinfo{person}{Rifat Shahriyar}, {and} \bibinfo{person}{Partha
  Chakroborty}.} \bibinfo{year}{2019}\natexlab{}.
\newblock \showarticletitle{Understanding the motivations, challenges and needs
  of blockchain software developers: A survey}.
\newblock \bibinfo{journal}{\emph{Empirical Software Engineering}}
  \bibinfo{volume}{24} (\bibinfo{year}{2019}), \bibinfo{pages}{2636--2673}.
\newblock
\urldef\tempurl%
\url{https://doi.org/10.1007/s10664-019-09708-7}
\showDOI{\tempurl}


\bibitem[Chen et~al\mbox{.}(2008)]%
        {ChenF2008}
\bibfield{author}{\bibinfo{person}{Feinian Chen}, \bibinfo{person}{Patrick~J.
  Curran}, \bibinfo{person}{Kenneth~A. Bollen}, \bibinfo{person}{James Kirby},
  {and} \bibinfo{person}{Pamela Paxton}.} \bibinfo{year}{2008}\natexlab{}.
\newblock \showarticletitle{An empirical evaluation of the use of fixed cutoff
  points in RMSEA test statistic in structural equation models}.
\newblock \bibinfo{journal}{\emph{Sociological Methods and Research}}
  \bibinfo{volume}{36} (\bibinfo{date}{5} \bibinfo{year}{2008}),
  \bibinfo{pages}{462--494}.
\newblock
Issue 4.
\showISSN{00491241}
\urldef\tempurl%
\url{https://doi.org/10.1177/0049124108314720}
\showDOI{\tempurl}


\bibitem[Chengalur-Smith et~al\mbox{.}(2010)]%
        {Chengalur2010}
\bibfield{author}{\bibinfo{person}{Indushobha Chengalur-Smith},
  \bibinfo{person}{Anna Sidorova}, {and} \bibinfo{person}{Sherae~L. Daniel}.}
  \bibinfo{year}{2010}\natexlab{}.
\newblock \showarticletitle{Sustainability of Free/Libre Open Source Projects:
  A Longitudinal Study}.
\newblock \bibinfo{journal}{\emph{Journal of the Association for Information
  Systems}}  \bibinfo{volume}{11} (\bibinfo{year}{2010}),
  \bibinfo{pages}{657--683}.
\newblock
Issue 11.
\showISSN{1536-9323}
\urldef\tempurl%
\url{https://doi.org/10.17705/1jais.00244}
\showDOI{\tempurl}


\bibitem[Clark and Bowles(2018)]%
        {ClarkD2018}
\bibfield{author}{\bibinfo{person}{{D. Angus} Clark} {and}
  \bibinfo{person}{Ryan~P. Bowles}.} \bibinfo{year}{2018}\natexlab{}.
\newblock \showarticletitle{Model Fit and Item Factor Analysis: Overfactoring,
  Underfactoring, and a Program to Guide Interpretation}.
\newblock \bibinfo{journal}{\emph{Multivariate Behavioral Research}}
  \bibinfo{volume}{53} (\bibinfo{date}{7} \bibinfo{year}{2018}),
  \bibinfo{pages}{544--558}.
\newblock
Issue 4.
\showISSN{00273171}
\urldef\tempurl%
\url{https://doi.org/10.1080/00273171.2018.1461058}
\showDOI{\tempurl}


\bibitem[Clark(2018)]%
        {Clark2018}
\bibfield{author}{\bibinfo{person}{Michael Clark}.}
  \bibinfo{year}{2018}\natexlab{}.
\newblock \bibinfo{booktitle}{\emph{Graphical \& Latent Variable Modeling}}.
\newblock
\urldef\tempurl%
\url{https://m-clark.github.io/sem/}
\showURL{%
Retrieved April 11, 2023 from \tempurl}


\bibitem[Costanza(1992)]%
        {Costanza1992}
\bibfield{author}{\bibinfo{person}{Robert Costanza}.}
  \bibinfo{year}{1992}\natexlab{}.
\newblock \bibinfo{booktitle}{\emph{Ecosystem health: New Goals for
  Environmental Management}}.
\newblock \bibinfo{publisher}{Island Press}, Chapter 14: Toward an operational
  definition of ecosystem health, \bibinfo{pages}{239--256}.
\newblock


\bibitem[Cronbach(1951)]%
        {Cronbach1951}
\bibfield{author}{\bibinfo{person}{Lee~J. Cronbach}.}
  \bibinfo{year}{1951}\natexlab{}.
\newblock \showarticletitle{Coefficient alpha and the internal structure of
  tests}.
\newblock \bibinfo{journal}{\emph{Psychometrika}}  \bibinfo{volume}{16}
  (\bibinfo{year}{1951}), \bibinfo{pages}{297--334}.
\newblock
\showISSN{1860-0980(Electronic),0033-3123(Print)}
\urldef\tempurl%
\url{https://doi.org/10.1007/BF02310555}
\showDOI{\tempurl}


\bibitem[Crowston et~al\mbox{.}(2006)]%
        {Crowston2006}
\bibfield{author}{\bibinfo{person}{Kevin Crowston}, \bibinfo{person}{James
  Howison}, {and} \bibinfo{person}{Hala Annabi}.}
  \bibinfo{year}{2006}\natexlab{}.
\newblock \showarticletitle{Information systems success in free and open source
  software development: Theory and measures}.
\newblock \bibinfo{journal}{\emph{Software Process Improvement and Practice}}
  \bibinfo{volume}{11} (\bibinfo{year}{2006}), \bibinfo{pages}{123--148}.
\newblock
Issue 2.
\showISSN{10774866}
\urldef\tempurl%
\url{https://doi.org/10.1002/spip.259}
\showDOI{\tempurl}


\bibitem[Dhungana et~al\mbox{.}(2010)]%
        {Dhungana2010}
\bibfield{author}{\bibinfo{person}{Deepak Dhungana}, \bibinfo{person}{Iris
  Groher}, \bibinfo{person}{Elisabeth Schludermann}, {and}
  \bibinfo{person}{Stefan Biffl}.} \bibinfo{year}{2010}\natexlab{}.
\newblock \showarticletitle{Software Ecosystems vs. Natural Ecosystems:
  Learning from the Ingenious Mind of Nature}. In
  \bibinfo{booktitle}{\emph{Proceedings of the Fourth European Conference on
  Software Architecture: Companion Volume}} (New York, NY, USA).
  \bibinfo{publisher}{Association for Computing Machinery},
  \bibinfo{pages}{96--102}.
\newblock
\showISBNx{9781450301794}
\urldef\tempurl%
\url{https://doi.org/10.1145/1842752.1842777}
\showDOI{\tempurl}


\bibitem[Due{\~n}as et~al\mbox{.}(2018)]%
        {Duenas2018}
\bibfield{author}{\bibinfo{person}{Santiago Due{\~n}as},
  \bibinfo{person}{Valerio Cosentino}, \bibinfo{person}{Gregorio Robles}, {and}
  \bibinfo{person}{Jes{\'u}s~M. Gonz{\'a}lez-Barahona}.}
  \bibinfo{year}{2018}\natexlab{}.
\newblock \showarticletitle{Perceval: software project data at your will}. In
  \bibinfo{booktitle}{\emph{Proceedings of the 40th International Conference on
  Software Engineering: Companion Proceeedings}}. ACM, \bibinfo{pages}{1--4}.
\newblock
\urldef\tempurl%
\url{https://doi.org/10.1145/3183440.3183475}
\showDOI{\tempurl}


\bibitem[Fabrigar and Wegener(2012)]%
        {Fabrigar2012}
\bibfield{author}{\bibinfo{person}{Leandre Fabrigar} {and}
  \bibinfo{person}{Duane Wegener}.} \bibinfo{year}{2012}\natexlab{}.
\newblock \bibinfo{booktitle}{\emph{Exploratory Factor Analysis}}.
\newblock \bibinfo{publisher}{Oxford University Press}.
\newblock
\showISBNx{978-0-19-973417-7}
\urldef\tempurl%
\url{https://doi.org/10.1093/acprof:osobl/9780199734177.001.0001}
\showDOI{\tempurl}


\bibitem[Fang and Neufeld(2008)]%
        {Fang2008}
\bibfield{author}{\bibinfo{person}{Yulin Fang} {and} \bibinfo{person}{Derrick
  Neufeld}.} \bibinfo{year}{2008}\natexlab{}.
\newblock \showarticletitle{{Understanding sustained participation in open
  source software projects}}.
\newblock \bibinfo{journal}{\emph{Journal of Management Information Systems}}
  \bibinfo{volume}{25}, \bibinfo{number}{4} (\bibinfo{year}{2008}),
  \bibinfo{pages}{9--50}.
\newblock
\showISSN{07421222}
\urldef\tempurl%
\url{https://doi.org/10.2753/MIS0742-1222250401}
\showDOI{\tempurl}


\bibitem[Finch(2020)]%
        {Finch2020}
\bibfield{author}{\bibinfo{person}{William~Holmes Finch}.}
  \bibinfo{year}{2020}\natexlab{}.
\newblock \showarticletitle{{Using Fit Statistic Differences to Determine the
  Optimal Number of Factors to Retain in an Exploratory Factor Analysis}}.
\newblock \bibinfo{journal}{\emph{Educational and Psychological Measurement}}
  \bibinfo{volume}{80}, \bibinfo{number}{2} (\bibinfo{year}{2020}),
  \bibinfo{pages}{217--241}.
\newblock
\showISSN{15523888}
\urldef\tempurl%
\url{https://doi.org/10.1177/0013164419865769}
\showDOI{\tempurl}


\bibitem[Franco-Bedoya et~al\mbox{.}(2014)]%
        {Franco2014}
\bibfield{author}{\bibinfo{person}{Oscar Franco-Bedoya}, \bibinfo{person}{David
  Ameller}, \bibinfo{person}{Dolors Costal}, {and} \bibinfo{person}{Xavier
  Franch}.} \bibinfo{year}{2014}\natexlab{}.
\newblock \showarticletitle{QuESo: a Quality Model for Open Source Software
  Ecosystems}.
\newblock \bibinfo{journal}{\emph{9th International Conference on Software
  Engineering and Applications (ICSOFT-EA)}}, \bibinfo{pages}{209--221}.
\newblock
\urldef\tempurl%
\url{https://doi.org/10.5220/0004993702090221}
\showDOI{\tempurl}


\bibitem[Ghapanchi(2015)]%
        {Ghapanchi2015}
\bibfield{author}{\bibinfo{person}{Amir~H. Ghapanchi}.}
  \bibinfo{year}{2015}\natexlab{}.
\newblock \showarticletitle{Investigating the Interrelationships among Success
  Measures of Open Source Software Projects}.
\newblock \bibinfo{journal}{\emph{Journal of Organizational Computing and
  Electronic Commerce}}  \bibinfo{volume}{25} (\bibinfo{year}{2015}),
  \bibinfo{pages}{28--46}.
\newblock
Issue 1.
\urldef\tempurl%
\url{https://doi.org/10.1080/10919392.2015.990775}
\showDOI{\tempurl}


\bibitem[GitHub(2021)]%
        {GitHub2021}
\bibfield{author}{\bibinfo{person}{GitHub}.} \bibinfo{year}{2021}\natexlab{}.
\newblock \bibinfo{booktitle}{\emph{{The 2021 State of the Octoverse}}}.
\newblock
\urldef\tempurl%
\url{https://octoverse.github.com/}
\showURL{%
Retrieved April 11, 2023 from \tempurl}


\bibitem[Goeminne and Mens(2013)]%
        {Goeminne2013}
\bibfield{author}{\bibinfo{person}{Mathieu Goeminne} {and} \bibinfo{person}{Tom
  Mens}.} \bibinfo{year}{2013}\natexlab{}.
\newblock \showarticletitle{{Analyzing ecosystems for open source software
  developer communities}}.
\newblock In \bibinfo{booktitle}{\emph{Software Ecosystems: Analyzing and
  Managing Business Networks in the Software Industry}},
  \bibfield{editor}{\bibinfo{person}{Slinger Jansen}} (Ed.). Number 2013.
  Chapter~12, \bibinfo{pages}{247--275}.
\newblock
\showISBNx{9781781955628}
\urldef\tempurl%
\url{https://doi.org/10.4337/9781781955635.00021}
\showDOI{\tempurl}


\bibitem[Goggins et~al\mbox{.}(2021)]%
        {Goggins2021}
\bibfield{author}{\bibinfo{person}{Sean Goggins}, \bibinfo{person}{Kevin
  Lumbard}, {and} \bibinfo{person}{Matt Germonprez}.}
  \bibinfo{year}{2021}\natexlab{}.
\newblock \showarticletitle{Open source community health: Analytical metrics
  and their corresponding narratives}.
\newblock \bibinfo{journal}{\emph{Proceedings - 2021 IEEE/ACM 4th International
  Workshop on Software Health in Projects, Ecosystems and Communities, SoHeal
  2021}} (\bibinfo{date}{5} \bibinfo{year}{2021}), \bibinfo{pages}{25--33}.
\newblock
\showISBNx{9781665445573}
\urldef\tempurl%
\url{https://doi.org/10.1109/SOHEAL52568.2021.00010}
\showDOI{\tempurl}


\bibitem[Gonzalez-Barahona(2021)]%
        {Gonzalez2021oss}
\bibfield{author}{\bibinfo{person}{Jesus~M. Gonzalez-Barahona}.}
  \bibinfo{year}{2021}\natexlab{}.
\newblock \showarticletitle{A brief history of free, open source software and
  its communities}.
\newblock \bibinfo{journal}{\emph{Computer}}  \bibinfo{volume}{54}
  (\bibinfo{date}{2} \bibinfo{year}{2021}), \bibinfo{pages}{75--79}.
\newblock
Issue 2.
\showISSN{15580814}
\urldef\tempurl%
\url{https://doi.org/10.1109/MC.2020.3041887}
\showDOI{\tempurl}


\bibitem[Gousios et~al\mbox{.}(2016)]%
        {Gousios2016}
\bibfield{author}{\bibinfo{person}{Georgios Gousios},
  \bibinfo{person}{Margaret~Anne Storey}, {and} \bibinfo{person}{Alberto
  Bacchelli}.} \bibinfo{year}{2016}\natexlab{}.
\newblock \showarticletitle{{Work practices and challenges in pull-based
  development: The contributor's perspective}}.
\newblock \bibinfo{journal}{\emph{Proceedings - International Conference on
  Software Engineering}}  \bibinfo{volume}{14-22-May-} (\bibinfo{year}{2016}),
  \bibinfo{pages}{285--296}.
\newblock
\showISBNx{9781450339001}
\showISSN{02705257}
\urldef\tempurl%
\url{https://doi.org/10.1145/2884781.2884826}
\showDOI{\tempurl}


\bibitem[{Hair Jr.} et~al\mbox{.}(2014)]%
        {Hair2014}
\bibfield{author}{\bibinfo{person}{Joseph~F. {Hair Jr.}},
  \bibinfo{person}{William~C. Black}, \bibinfo{person}{Barry~J. Babin}, {and}
  \bibinfo{person}{Rolph~E. Anderson}.} \bibinfo{year}{2014}\natexlab{}.
\newblock \bibinfo{booktitle}{\emph{{Multivariate Data Analysis}}
  (\bibinfo{edition}{seventh} ed.)}.
\newblock \bibinfo{publisher}{Pearson Education Limited},
  \bibinfo{address}{Essex}.
\newblock
\showISBNx{978-1-292-02190-4}


\bibitem[Hars and Ou(2001)]%
        {Hars2001}
\bibfield{author}{\bibinfo{person}{Alexander Hars} {and}
  \bibinfo{person}{Shaosong Ou}.} \bibinfo{year}{2001}\natexlab{}.
\newblock \showarticletitle{Working for free? - Motivations of participating in
  open source projects}.
\newblock \bibinfo{journal}{\emph{Proceedings of the Hawaii International
  Conference on System Sciences}}, \bibinfo{pages}{163}.
\newblock
\showISSN{10603425}
\urldef\tempurl%
\url{https://doi.org/10.1109/hicss.2001.927045}
\showDOI{\tempurl}


\bibitem[Hartigh et~al\mbox{.}(2013)]%
        {Hartigh2013}
\bibfield{author}{\bibinfo{person}{Erik~Den Hartigh}, \bibinfo{person}{Wouter
  Visscher}, {and} \bibinfo{person}{Michiel Tol}.}
  \bibinfo{year}{2013}\natexlab{}.
\newblock \bibinfo{booktitle}{\emph{Measuring the health of a business
  ecosystem}}.
\newblock \bibinfo{publisher}{Edward Elgar Publishing Limited},
  \bibinfo{pages}{221--246}.
\newblock


\bibitem[Hata et~al\mbox{.}(2022)]%
        {Hata2022}
\bibfield{author}{\bibinfo{person}{Hideaki Hata}, \bibinfo{person}{Nicole
  Novielli}, \bibinfo{person}{Sebastian Baltes},
  \bibinfo{person}{Raula~Gaikovina Kula}, {and} \bibinfo{person}{Christoph
  Treude}.} \bibinfo{year}{2022}\natexlab{}.
\newblock \showarticletitle{{GitHub Discussions: An exploratory study of early
  adoption}}.
\newblock \bibinfo{journal}{\emph{Empirical Software Engineering}}
  \bibinfo{volume}{27}, \bibinfo{number}{1} (\bibinfo{year}{2022}),
  \bibinfo{pages}{1--32}.
\newblock
\showISSN{15737616}
\urldef\tempurl%
\url{https://doi.org/10.1007/s10664-021-10058-6}
\showDOI{\tempurl}


\bibitem[Hu et~al\mbox{.}(2016)]%
        {Hu2016}
\bibfield{author}{\bibinfo{person}{Yan Hu}, \bibinfo{person}{Jun Zhang},
  \bibinfo{person}{Xiaomei Bai}, \bibinfo{person}{Shuo Yu}, {and}
  \bibinfo{person}{Zhuo Yang}.} \bibinfo{year}{2016}\natexlab{}.
\newblock \showarticletitle{{Influence analysis of Github repositories}}.
\newblock \bibinfo{journal}{\emph{SpringerPlus}} \bibinfo{volume}{5},
  \bibinfo{number}{1} (\bibinfo{year}{2016}).
\newblock
\showISSN{21931801}
\urldef\tempurl%
\url{https://doi.org/10.1186/s40064-016-2897-7}
\showDOI{\tempurl}


\bibitem[Iansiti and Levien(2004)]%
        {Iansiti2004}
\bibfield{author}{\bibinfo{person}{Marco Iansiti} {and} \bibinfo{person}{Roy
  Levien}.} \bibinfo{year}{2004}\natexlab{}.
\newblock \showarticletitle{Strategy as Ecology}.
\newblock \bibinfo{journal}{\emph{Harvard Business Review}}
  \bibinfo{volume}{BR0403} (\bibinfo{year}{2004}), \bibinfo{pages}{68--78}.
\newblock
\showISSN{0017-8012}
\urldef\tempurl%
\url{https://hbr.org/2004/03/strategy-as-ecology}
\showURL{%
\tempurl}


\bibitem[Jansen(2014)]%
        {Jansen2014}
\bibfield{author}{\bibinfo{person}{Slinger Jansen}.}
  \bibinfo{year}{2014}\natexlab{}.
\newblock \showarticletitle{Measuring the health of open source software
  ecosystems: Beyond the scope of project health}.
\newblock \bibinfo{journal}{\emph{Information and Software Technology}}
  \bibinfo{volume}{56} (\bibinfo{year}{2014}), \bibinfo{pages}{1508--1519}.
\newblock
Issue 11.
\showISSN{09505849}
\urldef\tempurl%
\url{https://doi.org/10.1016/j.infsof.2014.04.006}
\showDOI{\tempurl}


\bibitem[Jansen et~al\mbox{.}(2009)]%
        {Jansen2009}
\bibfield{author}{\bibinfo{person}{Slinger Jansen}, \bibinfo{person}{Anthony
  Finkelstein}, {and} \bibinfo{person}{Sjaak Brinkkemper}.}
  \bibinfo{year}{2009}\natexlab{}.
\newblock \showarticletitle{A Sense of Community: A Research Agenda for
  Software Ecosystems}.
\newblock \bibinfo{journal}{\emph{31st International Conference on Software
  Engineering, New and Emerging Research Track}}, \bibinfo{pages}{187--190}.
\newblock
\urldef\tempurl%
\url{https://slingerjansen.files.wordpress.com/2009/04/ssnniericse.pdf}
\showURL{%
\tempurl}


\bibitem[Link and Germonprez(2018)]%
        {Link2018}
\bibfield{author}{\bibinfo{person}{Georg~J.P. Link} {and} \bibinfo{person}{Matt
  Germonprez}.} \bibinfo{year}{2018}\natexlab{}.
\newblock \showarticletitle{Assessing Open Source Project Health}. In
  \bibinfo{booktitle}{\emph{{AMCIS 2018 Proceedings}}}.
  \bibinfo{publisher}{Association for Information Systems}.
\newblock
\showISBNx{9780996683166}
\urldef\tempurl%
\url{https://aisel.aisnet.org/amcis2018/Openness/Presentations/5}
\showURL{%
\tempurl}


\bibitem[Manikas and Hansen(2013)]%
        {Manikas2013}
\bibfield{author}{\bibinfo{person}{Konstantinos Manikas} {and}
  \bibinfo{person}{Klaus~Marius Hansen}.} \bibinfo{year}{2013}\natexlab{}.
\newblock \showarticletitle{Reviewing the Health of Software Ecosystems - A
  Conceptual Framework Proposal}. In \bibinfo{booktitle}{\emph{{5th Workshop on
  Software Ecosystems (IWSECO)}}}.
\newblock


\bibitem[McDonald(1999)]%
        {Mcdonald1999}
\bibfield{author}{\bibinfo{person}{Roderick~P. McDonald}.}
  \bibinfo{year}{1999}\natexlab{}.
\newblock \bibinfo{booktitle}{\emph{Test Theory: A Unified Treatment}
  (\bibinfo{edition}{1st} ed.)}.
\newblock \bibinfo{publisher}{Psychology Press}, \bibinfo{address}{New York}.
\newblock
\showISBNx{9781410601087}
\urldef\tempurl%
\url{https://doi.org/10.4324/9781410601087}
\showDOI{\tempurl}


\bibitem[Milovidov(2020)]%
        {Milovidov2020}
\bibfield{author}{\bibinfo{person}{Alexey Milovidov}.}
  \bibinfo{year}{2020}\natexlab{}.
\newblock \bibinfo{booktitle}{\emph{{Everything You Ever Wanted To Know About
  GitHub (But Were Afraid To Ask)}}}.
\newblock
\urldef\tempurl%
\url{https://ghe.clickhouse.tech/}
\showURL{%
Retrieved April 11, 2023 from \tempurl}


\bibitem[Moore(1993)]%
        {Moore1993}
\bibfield{author}{\bibinfo{person}{James~F. Moore}.}
  \bibinfo{year}{1993}\natexlab{}.
\newblock \showarticletitle{Predators and Prey: A New Ecology of Competition.}
\newblock \bibinfo{journal}{\emph{Harvard Business Review}}
  \bibinfo{volume}{71} (\bibinfo{year}{1993}), \bibinfo{pages}{75 -- 86}.
\newblock
Issue 3.
\showISSN{00178012}


\bibitem[Nakamoto(2008)]%
        {Nakamoto2008}
\bibfield{author}{\bibinfo{person}{Satoshi Nakamoto}.}
  \bibinfo{year}{2008}\natexlab{}.
\newblock \bibinfo{booktitle}{\emph{Bitcoin: A Peer-to-Peer Electronic Cash
  System}}.
\newblock
\urldef\tempurl%
\url{https://bitcoin.org/bitcoin.pdf}
\showURL{%
Retrieved April 11, 2023 from \tempurl}


\bibitem[Naparat et~al\mbox{.}(2015)]%
        {Naparat2015}
\bibfield{author}{\bibinfo{person}{Damrongsak Naparat},
  \bibinfo{person}{Michael Cahalane}, {and} \bibinfo{person}{Patrick
  Finnegan}.} \bibinfo{year}{2015}\natexlab{}.
\newblock \bibinfo{title}{Healthy Community and Healthy Commons: `Opensourcing'
  as a Sustainable Model of Software Production}.
\newblock
\newblock
\urldef\tempurl%
\url{https://doi.org/10.3127/ajis.v19i0.1221}
\showDOI{\tempurl}


\bibitem[Negoita et~al\mbox{.}(2019)]%
        {Negoita2019}
\bibfield{author}{\bibinfo{person}{Bogdan Negoita}, \bibinfo{person}{Gregory
  Vial}, \bibinfo{person}{Maha Shaikh}, {and} \bibinfo{person}{Aur{\'e}lie
  Labbe}.} \bibinfo{year}{2019}\natexlab{}.
\newblock \showarticletitle{Code forking and software development project
  sustainability: Evidence from GitHub}.
\newblock \bibinfo{journal}{\emph{40th International Conference on Information
  Systems, ICIS 2019}}.
\newblock
\showISBNx{9780996683197}
\urldef\tempurl%
\url{https://aisel.aisnet.org/icis2019/is_development/is_development/7}
\showURL{%
\tempurl}


\bibitem[Nijsse and Litchfield(2023)]%
        {Nijsse2023}
\bibfield{author}{\bibinfo{person}{Jeff Nijsse} {and} \bibinfo{person}{Alan
  Litchfield}.} \bibinfo{year}{2023}\natexlab{}.
\newblock \showarticletitle{Identifying Developer Engagement in Open Source
  Software Blockchain Projects through Factor Analysis}.
\newblock \bibinfo{journal}{\emph{56th Hawaii International Conference on
  System Sciences}}, \bibinfo{pages}{5333--5342}.
\newblock
\showISBNx{978-0-9981331-6-4}
\urldef\tempurl%
\url{https://hdl.handle.net/10125/103285}
\showURL{%
\tempurl}


\bibitem[Osman and Baysal(2021)]%
        {Osman2021}
\bibfield{author}{\bibinfo{person}{Khadija Osman} {and} \bibinfo{person}{Olga
  Baysal}.} \bibinfo{year}{2021}\natexlab{}.
\newblock \showarticletitle{Health is Wealth: Evaluating the Health of the
  Bitcoin Ecosystem in GitHub}. In \bibinfo{booktitle}{\emph{{IEEE/ACM 4th
  International Workshop on Software Health in Projects, Ecosystems and
  Communities (SoHeal)}}}.
\newblock
\showISBNx{9781665445573}
\urldef\tempurl%
\url{https://doi.org/10.1109/SoHeal52568.2021.00007}
\showDOI{\tempurl}


\bibitem[Perens(1999)]%
        {Perens1999}
\bibfield{author}{\bibinfo{person}{Bruce Perens}.}
  \bibinfo{year}{1999}\natexlab{}.
\newblock \showarticletitle{{The Open Source Definition}}.
\newblock In \bibinfo{booktitle}{\emph{Open Sources: Voices from the Open
  Source Revolution}}, \bibfield{editor}{\bibinfo{person}{Chris DiBona},
  \bibinfo{person}{Sam Ockman}, {and} \bibinfo{person}{Mark Stone}} (Eds.).
  \bibinfo{publisher}{O'Reilly {\&} Associates}, \bibinfo{pages}{171--188}.
\newblock
\showISBNx{1-56592-582-3}


\bibitem[Poba-Nzaou and Uwizeyemungu(2019)]%
        {Poba2019}
\bibfield{author}{\bibinfo{person}{Placide Poba-Nzaou} {and}
  \bibinfo{person}{Sylvestre Uwizeyemungu}.} \bibinfo{year}{2019}\natexlab{}.
\newblock \showarticletitle{{Worries of open source projects' contributors:
  Patterns, structures and engagement implications}}.
\newblock \bibinfo{journal}{\emph{Computers in Human Behavior}}
  \bibinfo{volume}{96}, \bibinfo{number}{September 2018}
  (\bibinfo{year}{2019}), \bibinfo{pages}{174--185}.
\newblock
\showISSN{07475632}
\urldef\tempurl%
\url{https://doi.org/10.1016/j.chb.2019.02.005}
\showDOI{\tempurl}


\bibitem[Raja and Tretter(2012)]%
        {Raja2012}
\bibfield{author}{\bibinfo{person}{Uzma Raja} {and}
  \bibinfo{person}{Marietta~J. Tretter}.} \bibinfo{year}{2012}\natexlab{}.
\newblock \showarticletitle{{Defining and evaluating a measure of Open Source
  Project survivability}}.
\newblock \bibinfo{journal}{\emph{IEEE Transactions on Software Engineering}}
  \bibinfo{volume}{38}, \bibinfo{number}{1} (\bibinfo{year}{2012}),
  \bibinfo{pages}{163--174}.
\newblock
\showISSN{00985589}
\urldef\tempurl%
\url{https://doi.org/10.1109/TSE.2011.39}
\showDOI{\tempurl}


\bibitem[Rapoport(1983)]%
        {Rapoport1983}
\bibfield{author}{\bibinfo{person}{Anatol Rapoport}.}
  \bibinfo{year}{1983}\natexlab{}.
\newblock \showarticletitle{The metaphor in the language of science}.
\newblock \bibinfo{journal}{\emph{Semiotische: Berlichte}}
  \bibinfo{volume}{12} (\bibinfo{year}{1983}), \bibinfo{pages}{25--43}.
\newblock
Issue 13.


\bibitem[Rapport(1989)]%
        {Rapport1989}
\bibfield{author}{\bibinfo{person}{David~J. Rapport}.}
  \bibinfo{year}{1989}\natexlab{}.
\newblock \showarticletitle{What constitutes ecosystem health?}
\newblock \bibinfo{journal}{\emph{Perspectives in Biology and Medicine}}
  \bibinfo{volume}{33} (\bibinfo{year}{1989}), \bibinfo{pages}{120--132}.
\newblock
Issue 1.
\showISSN{00315982}
\urldef\tempurl%
\url{https://doi.org/10.1353/pbm.1990.0004}
\showDOI{\tempurl}


\bibitem[Rapport et~al\mbox{.}(1998)]%
        {Rapport1998}
\bibfield{author}{\bibinfo{person}{David~J. Rapport}, \bibinfo{person}{Robert
  Costanza}, {and} \bibinfo{person}{Anthony~J. McMichael}.}
  \bibinfo{year}{1998}\natexlab{}.
\newblock \showarticletitle{Assessing ecosystem health}.
\newblock \bibinfo{journal}{\emph{Trends in Ecology \& Evolution}}
  \bibinfo{volume}{13}, \bibinfo{number}{10} (\bibinfo{year}{1998}),
  \bibinfo{pages}{397--402}.
\newblock
\showISSN{0169-5347}
\urldef\tempurl%
\url{https://doi.org/10.1016/S0169-5347(98)01449-9}
\showDOI{\tempurl}


\bibitem[Robinson et~al\mbox{.}(2016)]%
        {Robinson2016}
\bibfield{author}{\bibinfo{person}{William~N. Robinson},
  \bibinfo{person}{Tianjie Deng}, {and} \bibinfo{person}{Zirun Qi}.}
  \bibinfo{year}{2016}\natexlab{}.
\newblock \showarticletitle{Developer behavior and sentiment from data mining
  open source repositories}.
\newblock \bibinfo{journal}{\emph{Proceedings of the Annual Hawaii
  International Conference on System Sciences}}  \bibinfo{volume}{2016-March},
  \bibinfo{pages}{3729--3738}.
\newblock
\showISBNx{9780769556703}
\urldef\tempurl%
\url{https://doi.org/10.1109/HICSS.2016.465}
\showDOI{\tempurl}


\bibitem[Robles et~al\mbox{.}(2005)]%
        {Robles2005}
\bibfield{author}{\bibinfo{person}{Gregorio Robles}, \bibinfo{person}{Juan~Jose
  Amor}, \bibinfo{person}{Jesus~M. Gonzalez-Barahona}, {and}
  \bibinfo{person}{Israel Herraiz}.} \bibinfo{year}{2005}\natexlab{}.
\newblock \showarticletitle{Evolution and growth in large libre software
  projects}. In \bibinfo{booktitle}{\emph{{International Workshop on Principles
  of Software Evolution (IWPSE)}}}, Vol.~\bibinfo{volume}{2005}.
  \bibinfo{pages}{165--174}.
\newblock
\showISBNx{0769523498}
\urldef\tempurl%
\url{https://doi.org/10.1109/IWPSE.2005.17}
\showDOI{\tempurl}


\bibitem[Saini et~al\mbox{.}(2020)]%
        {Saini2020}
\bibfield{author}{\bibinfo{person}{Munish Saini}, \bibinfo{person}{Rohan
  Verma}, \bibinfo{person}{Antarpuneet Singh}, {and} \bibinfo{person}{Kuljit~K.
  Chahal}.} \bibinfo{year}{2020}\natexlab{}.
\newblock \showarticletitle{Investigating diversity and impact of the
  popularity metrics for ranking software packages}.
\newblock \bibinfo{journal}{\emph{Journal of Software: Evolution and Process}}
  \bibinfo{volume}{32} (\bibinfo{year}{2020}).
\newblock
Issue 9.
\urldef\tempurl%
\url{https://doi.org/10.1002/smr.2265}
\showDOI{\tempurl}


\bibitem[Schaeffer et~al\mbox{.}(1988)]%
        {Schaeffer1988}
\bibfield{author}{\bibinfo{person}{David~J. Schaeffer},
  \bibinfo{person}{Edwin~E. Herricks}, {and} \bibinfo{person}{Harold~W.
  Kerster}.} \bibinfo{year}{1988}\natexlab{}.
\newblock \showarticletitle{Ecosystem Health: I. Measuring Ecosystem Health}.
\newblock \bibinfo{journal}{\emph{Environmental Management}}
  \bibinfo{volume}{12} (\bibinfo{year}{1988}), \bibinfo{pages}{445--455}.
\newblock
Issue 4.
\urldef\tempurl%
\url{https://doi.org/10.1007/BF01873258}
\showDOI{\tempurl}


\bibitem[Schroer and Hertel(2009)]%
        {Schroer2009}
\bibfield{author}{\bibinfo{person}{Joachim Schroer} {and}
  \bibinfo{person}{Guido Hertel}.} \bibinfo{year}{2009}\natexlab{}.
\newblock \showarticletitle{{Voluntary engagement in an open web-based
  encyclopedia: Wikipedians and why they do it}}.
\newblock \bibinfo{journal}{\emph{Media Psychology}} \bibinfo{volume}{12},
  \bibinfo{number}{1} (\bibinfo{year}{2009}), \bibinfo{pages}{96--120}.
\newblock
\showISBNx{1521326080266}
\showISSN{15213269}
\urldef\tempurl%
\url{https://doi.org/10.1080/15213260802669466}
\showDOI{\tempurl}


\bibitem[Schwartz(2022)]%
        {Schwartz2022}
\bibfield{author}{\bibinfo{person}{Leo Schwartz}.}
  \bibinfo{year}{2022}\natexlab{}.
\newblock \showarticletitle{The 5 biggest crypto hacks of 2022.}
\newblock \bibinfo{journal}{\emph{Fortune.com}} (\bibinfo{year}{2022}).
\newblock
\urldef\tempurl%
\url{https://fortune.com/crypto/2022/12/30/5-biggest-crypto-hacks-2022/}
\showURL{%
\tempurl}


\bibitem[Shaikh and Levina(2019)]%
        {Shaikh2019}
\bibfield{author}{\bibinfo{person}{Maha Shaikh} {and} \bibinfo{person}{Natalia
  Levina}.} \bibinfo{year}{2019}\natexlab{}.
\newblock \showarticletitle{{Selecting an open innovation community as an
  alliance partner: Looking for healthy communities and ecosystems}}.
\newblock \bibinfo{journal}{\emph{Research Policy}} \bibinfo{volume}{48},
  \bibinfo{number}{8} (\bibinfo{year}{2019}), \bibinfo{pages}{103766}.
\newblock
\showISSN{00487333}
\urldef\tempurl%
\url{https://doi.org/10.1016/j.respol.2019.03.011}
\showDOI{\tempurl}


\bibitem[Shi and Sun(2021)]%
        {Shi2021}
\bibfield{author}{\bibinfo{person}{Zhengzhong Shi} {and} \bibinfo{person}{Hua
  Sun}.} \bibinfo{year}{2021}\natexlab{}.
\newblock \showarticletitle{{Sustained Participation in Open Source Software
  Project Communities}}.
\newblock \bibinfo{journal}{\emph{Journal of Computer Information Systems}}
  \bibinfo{volume}{00}, \bibinfo{number}{00} (\bibinfo{year}{2021}),
  \bibinfo{pages}{1--14}.
\newblock
\showISSN{23802057}
\urldef\tempurl%
\url{https://doi.org/10.1080/08874417.2021.1949645}
\showDOI{\tempurl}


\bibitem[Smirnova et~al\mbox{.}(2022)]%
        {Smirnova2022}
\bibfield{author}{\bibinfo{person}{Inna Smirnova}, \bibinfo{person}{Markus
  Reitzig}, {and} \bibinfo{person}{Oliver Alexy}.}
  \bibinfo{year}{2022}\natexlab{}.
\newblock \showarticletitle{What makes the right OSS contributor tick?
  Treatments to motivate high-skilled developers}.
\newblock \bibinfo{journal}{\emph{Research Policy}}  \bibinfo{volume}{51}
  (\bibinfo{year}{2022}).
\newblock
Issue 1.
\urldef\tempurl%
\url{https://doi.org/10.1016/j.respol.2021.104368}
\showDOI{\tempurl}


\bibitem[Tamburri et~al\mbox{.}(2019)]%
        {Tamburri2019}
\bibfield{author}{\bibinfo{person}{Damian~A. Tamburri}, \bibinfo{person}{Fabio
  Palomba}, \bibinfo{person}{Alexander Serebrenik}, {and} \bibinfo{person}{Andy
  Zaidman}.} \bibinfo{year}{2019}\natexlab{}.
\newblock \bibinfo{booktitle}{\emph{{Discovering community patterns in
  open-source: a systematic approach and its evaluation}}}.
  Vol.~\bibinfo{volume}{24}.
\newblock \bibinfo{publisher}{Empirical Software Engineering}.
\newblock
\showISBNx{1066401896}
\showISSN{15737616}
\urldef\tempurl%
\url{https://doi.org/10.1007/s10664-018-9659-9}
\showDOI{\tempurl}


\bibitem[van~den Berk et~al\mbox{.}(2010)]%
        {vandenBerk2010}
\bibfield{author}{\bibinfo{person}{Ivo van~den Berk}, \bibinfo{person}{Slinger
  Jansen}, {and} \bibinfo{person}{L\'{u}tzen Luinenburg}.}
  \bibinfo{year}{2010}\natexlab{}.
\newblock \showarticletitle{Software Ecosystems: A Software Ecosystem Strategy
  Assessment Model}. In \bibinfo{booktitle}{\emph{Proceedings of the Fourth
  European Conference on Software Architecture: Companion Volume}} (Copenhagen,
  Denmark) \emph{(\bibinfo{series}{ECSA '10})}. \bibinfo{publisher}{Association
  for Computing Machinery}, \bibinfo{address}{New York, NY, USA},
  \bibinfo{pages}{127–134}.
\newblock
\showISBNx{9781450301794}
\urldef\tempurl%
\url{https://doi.org/10.1145/1842752.1842781}
\showDOI{\tempurl}


\bibitem[Wahyudin et~al\mbox{.}(2007)]%
        {Wahyudin2007}
\bibfield{author}{\bibinfo{person}{Dindin Wahyudin}, \bibinfo{person}{Khabib
  Mustofa}, \bibinfo{person}{Alexander Schatten}, \bibinfo{person}{Stefan
  Biffl}, {and} \bibinfo{person}{A~Min Tjoa}.} \bibinfo{year}{2007}\natexlab{}.
\newblock \showarticletitle{Monitoring the ``health'' status of open source
  web-engineering projects}.
\newblock \bibinfo{journal}{\emph{International Journal of Web Information
  Systems}}  \bibinfo{volume}{3} (\bibinfo{year}{2007}),
  \bibinfo{pages}{116--139}.
\newblock
Issue 1.
\showISSN{17440084}
\urldef\tempurl%
\url{https://doi.org/10.1108/17440080710829252}
\showDOI{\tempurl}


\bibitem[Wang et~al\mbox{.}(2020)]%
        {Wang2020}
\bibfield{author}{\bibinfo{person}{Zhendong Wang}, \bibinfo{person}{Yang Feng},
  \bibinfo{person}{Yi Wang}, \bibinfo{person}{James~A. Jones}, {and}
  \bibinfo{person}{David Redmiles}.} \bibinfo{year}{2020}\natexlab{}.
\newblock \showarticletitle{{Unveiling Elite Developers Activities in Open
  Source Projects}}.
\newblock \bibinfo{journal}{\emph{ACM Transactions on Software Engineering and
  Methodology}} \bibinfo{volume}{29}, \bibinfo{number}{3}
  (\bibinfo{year}{2020}).
\newblock
\showISSN{15577392}
\urldef\tempurl%
\url{https://doi.org/10.1145/3387111}
\showDOI{\tempurl}


\bibitem[Wang and Perry(2016)]%
        {Wang2016}
\bibfield{author}{\bibinfo{person}{Zhongjie Wang} {and}
  \bibinfo{person}{Dewayne~E. Perry}.} \bibinfo{year}{2016}\natexlab{}.
\newblock \showarticletitle{Role distribution and transformation in open source
  software project teams}.
\newblock \bibinfo{journal}{\emph{Proceedings - Asia-Pacific Software
  Engineering Conference, APSEC}}  \bibinfo{volume}{2016-May},
  \bibinfo{pages}{119--126}.
\newblock
\showISBNx{9781467396448}
\urldef\tempurl%
\url{https://doi.org/10.1109/APSEC.2015.12}
\showDOI{\tempurl}


\bibitem[Watkins(2018)]%
        {Watkins2018}
\bibfield{author}{\bibinfo{person}{Marley~W. Watkins}.}
  \bibinfo{year}{2018}\natexlab{}.
\newblock \showarticletitle{{Exploratory Factor Analysis: A Guide to Best
  Practice}}.
\newblock \bibinfo{journal}{\emph{Journal of Black Psychology}}
  \bibinfo{volume}{44}, \bibinfo{number}{3} (\bibinfo{year}{2018}),
  \bibinfo{pages}{219--246}.
\newblock
\showISSN{15524558}
\urldef\tempurl%
\url{https://doi.org/10.1177/0095798418771807}
\showDOI{\tempurl}


\end{thebibliography}
\end{document}